\documentclass[aps,showpacs,twocolumn,longbibliography]{revtex4-1}
\usepackage[utf8]{inputenc}
\usepackage{amsfonts}
\usepackage{amssymb}
\usepackage{amsmath}
\usepackage{graphicx}
\usepackage{epsfig}
\usepackage{subfigure}
\usepackage{appendix}
\usepackage{color}
\usepackage{hyperref}
\usepackage{clrscode3e}

\hypersetup{hypertex=true,
	colorlinks=true,
	linkcolor=blue,
	anchorcolor=blue,
	urlcolor=blue,
	citecolor=blue}

\begin{document}
	
	\title{Extended Wannier-Stark ladder and particle-pair Bloch oscillations in
		dimerized non-Hermitian systems}
	\author{H. P. Zhang}
	\author{Z. Song}
	\email{songtc@nankai.edu.cn}
	
	\begin{abstract}
		In the Hermitian regime, the Wannier-Stark ladder characterizes the
		eigenstates of an electron in a periodic potential with an applied static
		electric field. In this work, we extend this concept to the complex regime
		for a periodic non-Hermitian system under a linear potential. We show that
		although the energy levels can be complex, they are still equally spaced by
		a real Bloch frequency. This ensures single-particle Bloch oscillations with
		a damping (or growing) rate. The system can also support standard
		two-particle Bloch oscillations under certain conditions. We propose two
		types of dimerized non-Hermitian systems to demonstrate our results. In
		addition, we also propose a scheme to demonstrate the results of
		particle-pair dynamics in a single-particle 2D $\mathcal{PT}$-symmetric
		square lattice.
	\end{abstract}
	
	\affiliation{School of Physics, Nankai University, Tianjin 300071, China}
	
	\maketitle
	
	\section{Introduction}
	
	\label{Introduction} Bloch oscillation is a feature of the dynamics of an
	electron under a periodic potential and an external electric field \cite%
	{Bloch1929}. Within the Hermitian regime, Bloch oscillations occur due to
	the acceleration of an electron by an electric field, which is described by
	the acceleration theorem in reciprocal space \cite{wannier1959elements}, and
	subsequent Bragg reflections from the periodic lattice potential at the
	boundaries of the first Brillouin zone. An alternative description can be
	given\ in the framework of traditional quantum mechanics. In 1960, Wannier \cite{Wannier1960} theoretically proved that the eigenstates of an electron
	in a periodic potential with an applied static electric field can be
	described by the Wannier Stark ladder. Then, the periodic dynamics arise
	from the Wannier-Stark ladder, which consists of quantized equidistant
	energy levels separated by the Bloch frequency \cite{Glueck2002}. As a
	universal wave phenomenon, it has attracted much attention from various
	research areas due to experimental observations. It has been reported that
	such periodic dynamics are observed in a semiconductor superlattice \cite%
	{Waschke1993}, ultracold atoms in the optical lattice \cite%
	{BenDahan1996,Wilkinson1996,Anderson1998,Morsch2001}, and many other systems
	sequentially \cite%
	{Morandotti1999,SanchisAlepuz2007,Meinert2017,Zhang2022,Hansen2022}.
	
	Theoretically, the investigation of the Wannier-Stark ladder has been
	extended to new frontiers, such as non-Hermitian systems \cite%
	{Longhi2009,Longhi2014}. Recent works \cite{Longhi2015,Graefe2016,Longhi2017}
	show that a non-Hermitian hopping term, including asymmetrical and complex
	strengths, remains the reality of the Wannier-Stark ladder. {
		A non-Hermitian system can exhibit exclusive dynamics that never
		occurs in a Hermitian system, especially when complex energy is involved.
		Importantly, non-Hermitian dynamics can be observed in experiments. In
		contrast to non-Hermitian optical systems, a few-body non-Hermitian
		Hamiltonian can be implemented through a scheme that is analogous to
		heralded entanglement protocols {\cite{Lee2014}.}} This raises the question
	of whether an extended Wannier-Stark ladder, which consists of quantized
	equidistant \textit{complex} energy levels, but separated by the \textit{real%
	} Bloch frequency, can emerge in a periodic non-Hermitian system under a
	linear potential.
	
	In this work, we focus on the effects of a linear potential on periodic
	systems, using the tight-binding model as a framework for analysis. This
	model is widely used in condensed matter physics to describe the behavior of
	electrons in crystalline structures. We extend the concept of the
	Wannier-Stark ladder, which is traditionally used to describe eigenstates in
	a periodic potential with an electric field in Hermitian systems, to
	non-Hermitian systems with complex eigenvalues. This extension is
	significant because it broadens the analytical tools available for studying
	the dynamics of such systems. Such an extension of the Wannier-Stark ladder
	to the complex regime enables the analytical prediction of the dynamics of
	many non-Hermitian systems with a fixed number of particles. This provides
	the prediction of Bloch oscillations in Non-Hermitian systems without the
	need for complex numerical simulations. To demonstrate this point, two types
	of dimerized non-Hermitian systems are proposed as examples to illustrate
	the research findings. This indicates that dimerized non-Hermitian systems
	can support both single-particle and two-particle Bloch oscillations. These
	oscillations are characterized by a damping (or growing) rate, which
	reflects the temporal evolution of the quantum state. Furthermore, we show
	that the extended Wannier-Stark ladder concept, initially developed for 1D
	systems, can also be applied to 2D systems. This broadens the applicability
	of the concept and opens up new avenues for related research. We
	propose a scheme involving electron-, boson-, and spinless fermion-pair
	dynamics in a single-particle 2D $\mathcal{PT}$-symmetric square lattice to
	demonstrate the results. Here, $\mathcal{PT}$ symmetry \cite%
	{Bender1998,Mostafazadeh2003,Mostafazadeh2004,Jones2005,Bender1999,Dorey2001,Bender2002,Bender2007}%
	, a combination of parity and time-reversal symmetry, is a key concept in
	non-Hermitian physics \cite{Ashida2020} and can lead to real eigenvalues
	under certain conditions, especially for descrete systems \cite%
	{Jin2010,Zhang2012}. Our findings pave the way for further investigations
	into a wide variety of periodic systems under the influence of a linear
	potential. This could have significant implications for understanding and
	designing devices with tailored properties. In addition, this work
	contributes to the understanding of non-Hermitian systems and provides new
	tools for their analysis.
	
	This paper is organized as follows. In Sec. \ref{General formalism}, we
	present a general formalism for Hamiltonians with ramped translational
	symmetry and show the existence of energy ladders, referred to as extended
	Wannier-Stark ladders, regardless of the Hermiticity of the systems. In Sec. %
	\ref{Dimerized non-Hermitian chain}, we propose two types of non-Hermitian
	dimer systems to demonstrate the extended Wannier-Stark ladders. In Sec. \ref%
	{Bloch oscillations}, we study single-particle damping Bloch oscillations.
	Sec. \ref{2D simulator for electron-pair dynamics} is devoted to
	the 2D representation of two-electron, two-boson, and two-spinless fermion
	dynamics in a 1D system.\ Finally, we summarize our results in Sec. \ref%
	{Summary}.
	
	\begin{figure}[t]
		\centering
		\includegraphics[width=0.4\textwidth]{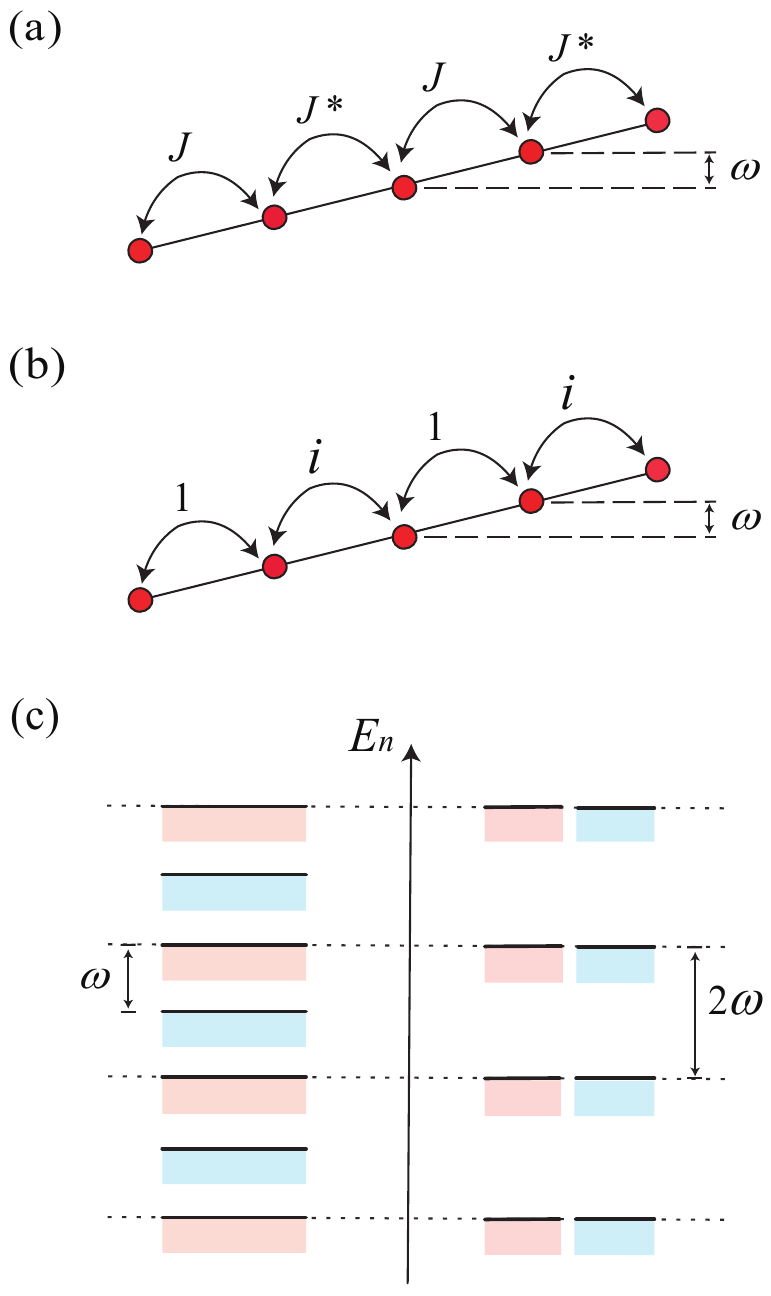}
		\caption{Schematic illustrations of (a) Hamiltonian in Eq.\ (\protect\ref%
			{J-J*}) and (b) Hamiltonian in Eq.\ (\protect\ref{1-i}), which correspond to
			the adjacent hopping strengths $J$, $J^{\ast }$ and $1$, $i$, respectively.
			The on-site potential is tilted with slope $\protect\omega $. It is shown
			that the latter has $\mathcal{PT}$ symmetry, being a pseudo-Hermitian
			system. (c) Energy level structure diagrams for Hamiltonian of (a) and (b).
			The solid line represents the real part of the energy level with
			isoenergetic distance $\protect\omega $ for (a) and $2\protect\omega $ for
			(b). The red and blue blocks represent the positive imaginary part and
			negative imaginary part of the energy level, respectively. It can be seen
			that the energy spectrum on the right side is composed of conjugate energy
			levels, which correspond exactly to the $\mathcal{PT}$ symmetry of (b).}
		\label{fig1}
	\end{figure}
	
	\section{General formalism}
	
	\label{General formalism}
	
	In this section, we present a general formalism\ supporting the extended
	Wannier-Stark ladder. First, we propose a class of Hamiltonians, including
	non-Hermitian ones, which possesses a ramped translational symmetry. Second,
	we show that such Hamiltonians can have several sets of energy ladders: the
	energy levels can be complex, but the level spacing can be real-valued and
	identical. The translational operators across two neighboring unit cells act
	as ladder operators, generating a set of eigenstates from an obtained
	eigenstate.
	
	\subsection{Model and symmetry}
	
	\label{Model and symmetry}
	
	We start with a general tight-binding model with the Hamiltonian in the form
	
	\begin{eqnarray}
		H &=&H_{0}+\omega \sum_{j=-\infty }^{\infty }ja_{j}^{\dag }a_{j},  \label{H}
		\\
		H_{0} &=&\sum_{i,j=-\infty }^{\infty }J_{ij}a_{i}^{\dag }a_{j},
	\end{eqnarray}%
	where $a_{i}^{\dag }$\ ($a_{i}$) is the boson or fermion creation
	(annihilation) operator at the $i$th site. It is an infinite chain lattice
	with a unit cell consisting of $n_{0}$ different types of sites. This means
	that the Hamiltonian $H_{0}$\ obeys the translational symmetry 
	\begin{equation}
		T_{n_{0}}H_{0}T_{n_{0}}^{-1}=H_{0},
	\end{equation}%
	where $T_{n_{0}}$\ is the translational operator defined as 
	\begin{equation}
		T_{n_{0}}a_{j}^{\dag }T_{n_{0}}^{-1}=a_{j+n_{0}}^{\dag
		},T_{n_{0}}a_{j}T_{n_{0}}^{-1}=a_{j+n_{0}}.
	\end{equation}%
	In addition, there are no other restrictions on the set of coefficients $%
	\left\{ J_{ij}\right\} $. Fig. \ref{fig1} is a schematic of the system. In
	this sense, the following conclusion still holds for the non-Hermitian
	Hamiltonian $H_{0}$. The non-Hermiticity arises from the case with $%
	J_{ij}\neq \left( J_{ji}\right) ^{\ast }$, which corresponds to
	non-Hermitian hopping strength for $i\neq j$, and complex on-site potentials
	for $i=j$. {In previous work \cite{ZHP2024}, a simple case with }$%
	J_{j(j+1)}=J_{(j+1)j}=J${\ (}$n_{0}=1$) is considered for complex $J$. {The
		exact solution indicates that the energy levels are always real and
		equidistant, in accordance with the above conclusion. }In general, the
	Hamiltonian $H_{0}$ can be block diagonalized due to translational symmetry,
	while it is difficult to diagonalize the Hamiltonian $H$\ with $n_{0}>1$,
	since term\ $\omega \sum_{j=-\infty }^{\infty }ja_{j}^{\dag }a_{j}$ breaks
	the translational symmetry.
	
	\subsection{Ladder operators}
	
	\label{Ladder operators}
	
	However, the special structure of the linear potential ensures that $H$\
	obeys 
	\begin{equation}
		T_{n_{0}}HT_{n_{0}}^{-1}=H-n_{0}\omega ,
	\end{equation}
	which is referred to as a ramped translational symmetry. Although this
	symmetry has no help in determining the explicit form of eigenstates, it
	reflects the relationships between the eigenstates.
	
	Suppose we have a solution $|\psi _{0}\rangle $ of the Schrodinger equation
	corresponding to energy $E_{0}$ 
	\begin{equation}
		H|\psi _{0}\rangle =E_{0}|\psi _{0}\rangle ,
	\end{equation}%
	we always have 
	\begin{equation}
		H\left( T_{n_{0}}|\psi _{0}\rangle \right) =\left( E_{0}+n_{0}\mathcal{\
			\omega }\right) \left( T_{n_{0}}|\psi _{0}\rangle \right) ,
	\end{equation}%
	and 
	\begin{equation}
		H\left( T_{n_{0}}^{-1}|\psi _{0}\rangle \right) =\left( E_{0}-n_{0}\mathcal{%
			\ \omega }\right) \left( T_{n_{0}}^{-1}|\psi _{0}\rangle \right) ,
	\end{equation}%
	i.e., $T_{n_{0}}|\psi _{0}\rangle $($T_{n_{0}}^{-1}|\psi _{0}\rangle $)\ is
	also the eigenstate of $H$\ with eigen energy $E_{0}+\mathcal{\omega }$($%
	E_{0}-\mathcal{\omega }$). Operator $T_{n_{0}}$\ moves up the energy ladder
	by a step of $n_{0}\mathcal{\omega }$ and the operator $T_{n_{0}}^{-1}$
	moves down the energy ladder by a step of $n_{0}\mathcal{\omega }$.\ We can
	then construct a set of eigenstates 
	\begin{equation}
		|\psi _{n}\rangle =\left( T_{n_{0}}\right) ^{n}|\psi _{0}\rangle ,
	\end{equation}%
	$(n=0,\pm 1,\pm 2,...)$ with eigenenergy 
	\begin{equation}
		E_{n}=E_{0}+nn_{0}\mathcal{\omega },
	\end{equation}%
	\begin{figure*}[tbh]
		\centering
		\includegraphics[width=0.9\textwidth]{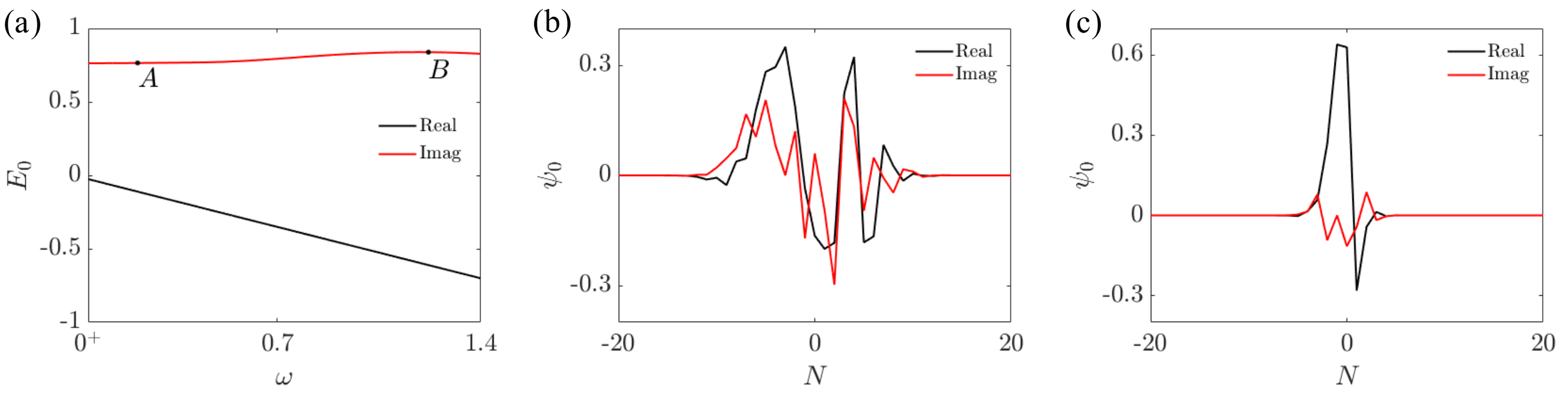}  
		\caption{Plots of $E_{0}$ and $\protect\psi _{0}$ defined in
			Eqs.\ (\protect\ref{El+-}) and (\protect\ref{e19}), respectively, for the
			Hamiltonian system (\protect\ref{1-i}) in a potential with slope $\protect%
			\omega $. (a) is the dependence curve of $E_{0}$ on $\protect\omega $. (b)
			and (c) are the eigenfunctions for the systems at points $A$ and $B$, where $%
			A$ and $B$ correspond to $\protect\omega =0.2$ and $1.2$, respectively. It
			can be observed that when $\protect\omega $ is small, the corresponding
			eigenfunction is wider, which corresponds to the pronounced Bloch
			oscillations that follow. When $\protect\omega $ is larger, the
			eigenfunction noticeably narrows, which is not conducive to observing Bloch
			oscillations. {The real part of }$E_{0}${\ is a linear
				function of }$\protect\omega $\textbf{.}}
		\label{fig2}
	\end{figure*}
	\begin{figure*}[tbh]
		\centering
		\includegraphics[width=0.9\textwidth]{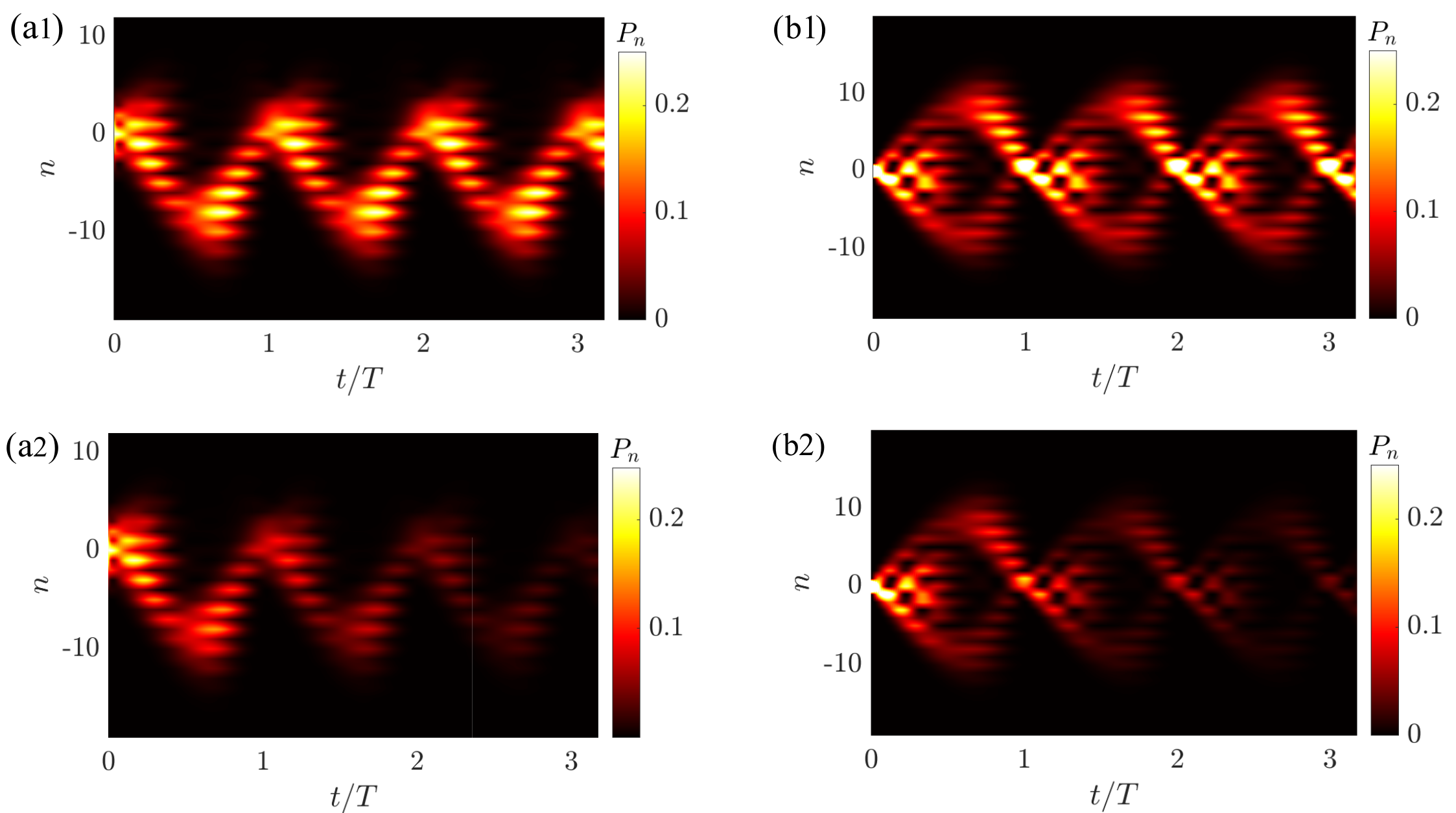}  
		\caption{Plots of $P_{n}(t)$ defined in Eq.\ (\protect\ref{Pj(t)}), obtained
			by numerical diagonalization for several different initial states and
			prefactors with $\protect\omega =0.2$, $T=\protect\pi /\protect\omega $.
			Here, (a1) and (a2) correspond to the initial state $e^{-\protect\alpha %
				^{2}n^{2}}\left( \protect\alpha =0.3\right) $ and the prefactors $\protect%
			\lambda =0.764$ and $0.787$, respectively. (b1) and (b2) correspond to the
			initial state $\protect\delta _{n0}$ and the prefactors $\protect\lambda %
			=0.764$ and $0.787$, respectively. }
		\label{fig3}
	\end{figure*}
	Thus, we can conclude that the spectrum of $H$ has equal-spacing energy
	levels. The translational operator $T_{n_{0}}$\ acts as a ladder operator.
	This proof is independent of $\left\{ J_{ij}\right\} $, i.e., it can be a
	complex number. We note that based on another eigenstate $|\phi _{0}\rangle $
	, another set of eigenstates can be generated accordingly.
	
	Applying the conclusion to the case with $J_{j(j+1)}=J_{(j+1)j}=J$\ ($%
	n_{0}=1 $), the same result can be obtained as that in Ref. \cite%
	{ZHP2024,ZKL2024} for complex constant $J$. In the following, we focus on
	two types of non-Hermitian systems to demonstrate the application of our
	conclusion and reveal the pair dynamics.
	
	\section{Dimerized non-Hermitian chains}
	
	\label{Dimerized non-Hermitian chain}In the following, we focus on
	non-Hermitian systems for two cases with $n_{0}=2$. Fig. \ref{fig1} is a
	schematics for the structure and energy levels of\ two systems.
	
	(i) $J_{2j(2j+1)}=J_{(2j+1)2j}=J$ and $J_{2j(2j-1)}=J_{(2j-1)2j}=J^{\ast }$;
	in this case, the Hamiltonian reads 
	\begin{equation}
		H_{0}=\sum_{l=-\infty }^{\infty }[J\left( a_{2l}^{\dag }a_{2l+1}+\mathrm{H.c.%
		}\right) +J^{\ast }\left( a_{2l-1}^{\dag }a_{2l}+\mathrm{H.c.}\right) ],
		\label{J-J*}
	\end{equation}%
	Fig. \ref{fig1}(a) is a schematics of the system. A straightforward
	derivation shows that%
	\begin{equation}
		\left\{ 
		\begin{array}{c}
			T_{2}HT_{2}^{-1}=H-2\omega \\ 
			T_{1}HT_{1}^{-1}=H^{\ast }-\omega%
		\end{array}%
		\right. .
	\end{equation}
	Starting from an arbitrary eigenstate $|\psi _{0}\rangle $ of $H$, a set of
	eigenstates can be generated as 
	\begin{equation}
		|\psi _{n}\rangle =\left\{ 
		\begin{array}{cc}
			\left( T_{2}\right) ^{l}|\psi _{0}\rangle , & n=2l \\ 
			\mathcal{T}\left( T_{1}\right) ^{2l+1}|\psi _{0}\rangle , & n=2l+1%
		\end{array}%
		\right.
	\end{equation}%
	$(l=0,\pm 1,\pm 2,...)$ with eigenenergy 
	\begin{equation}
		E_{n}=\left\{ 
		\begin{array}{cc}
			E_{0}+n\mathcal{\omega }, & n=2l \\ 
			E_{0}^{\ast }+n\mathcal{\omega }, & n=2l+1%
		\end{array}%
		\right. ,
	\end{equation}%
	or in the compact form 
	\begin{equation}
		|\psi _{n}\rangle =\left( \mathcal{T}T_{1}\right) ^{n}|\psi _{0}\rangle
	\end{equation}%
	with 
	\begin{equation}
		E_{n}=\left( \mathcal{T}\right) ^{n}E_{0}\left( \mathcal{T}^{-1}\right)
		^{n}+n\mathcal{\omega },
	\end{equation}%
	$(n=0,\pm 1,\pm 2,...)$, where $\mathcal{T}$\ is the time-reversal operator
	defined as $\mathcal{T}\sqrt{-1}\mathcal{T}^{-1}$ $=-\sqrt{-1}$. We can see
	that operator $\mathcal{T}T_{1}$\ acts as a ladder operator. Considering the
	eigenstates in a single-particle invariant subspace, complete eigenstates
	can be constructed from an arbitrary eigenstate. The corresponding energy
	levels are complex: the real part has equal spacing, while the imaginary
	part is alternatively conjugated.
	
	(ii) $J_{2j(2j+1)}=J_{(2j+1)2j}=1$ and $J_{2j(2j-1)}=J_{(2j-1)2j}=i$; in
	this case, the Hamiltonian reads
	
	\begin{equation}
		H_{0}=\sum_{l=-\infty }^{\infty }[\left( a_{2l}^{\dag }a_{2l+1}+\mathrm{H.c.}%
		\right) +i\left( a_{2l-1}^{\dag }a_{2l}+\mathrm{H.c.}\right) ],  \label{1-i}
	\end{equation}%
	Fig. \ref{fig1}(b) is a schematics of the system. A straightforward
	derivation shows that%
	\begin{equation}
		\left\{ 
		\begin{array}{c}
			T_{2}HT_{2}^{-1}=H-2\omega  \\ 
			gHg^{-1}=H^{\ast }%
		\end{array}%
		\right. ,
	\end{equation}%
	where the local gauge transformation is defined as 
	\begin{equation}
		\left\{ 
		\begin{array}{c}
			ga_{2l}g^{-1}=(-1)^{l}a_{2l} \\ 
			ga_{2l+1}g^{-1}=(-1)^{l}a_{2l+1}%
		\end{array}%
		\right. .
	\end{equation}%
	Starting from an arbitrary eigenstate $|\psi _{0}\rangle $ of $H$, a set of
	eigenstates can be generated as 
	\begin{equation}
		\left\{ 
		\begin{array}{c}
			|\psi _{l}^{+}\rangle =\left( T_{2}\right) ^{l}|\psi _{0}\rangle  \\ 
			|\psi _{l}^{-}\rangle =\mathcal{T}g\left( T_{2}\right) ^{l}|\psi _{0}\rangle 
		\end{array}%
		\right. ,  \label{e19}
	\end{equation}%
	$(l=0,\pm 1,\pm 2,...)$ with eigenenergy%
	\begin{equation}
		\left\{ 
		\begin{array}{c}
			E_{l}^{+}=E_{0}+2l\mathcal{\omega } \\ 
			E_{l}^{-}=E_{0}^{\ast }+2l\mathcal{\omega }%
		\end{array}%
		\right. .  \label{El+-}
	\end{equation}%
	Here, without loss of generality, we assume that $\func{Im}E_{0}>0$.
	Considering the eigenstates in a single-particle invariant subspace,
	complete eigenstates can be constructed from an arbitrary eigenstate. The
	spectrum indicates that the Hamiltonian is pseudo Hermitian, in which
	complex energy levels appear in the conjugate pair. { 
		This feature allows to construct a two-particle subspace with all real
		energy levels. Considering a set of two-particle eigenstates $|\psi
		_{l}^{+}\rangle |\psi _{l}^{-}\rangle ${, including electron, boson and
			spinless fermion, the corrsponding energy levels are\ }$2\func{Re}\left(
		E_{0}\right) +4l\mathcal{\omega }${, forming a standard Wannier-Stark
			ladder. Additionally, as a starting point, }$E_{0}${\ and }$|\psi
		_{0}\rangle ${\ can obtained numerically for an infinite system by using the
			powerful Floquet operator method  \cite{Maksimov2015,Maksimov2015a}. }In
		Fig.\ \ref{fig2}, the plots of $E_{0}$ as a function of $\omega $ and the
		profiles of two typical $|\psi _{0}\rangle $\ states are presented. We find
		that the real part of $E_{0}$ is a linear function of $\omega $ and the
		width of $|\psi _{0}\rangle $\ strongly depends on the value of $\omega $.}
	
	\section{Bloch oscillations}
	
	\label{Bloch oscillations}
	
	In this section, we investigate the dynamic feature of a system with
	extended Wannier-Stark ladder. We will consider the single- and
	double-particle dynamics driven by the Hamiltonian in Eq. (\ref{1-i}) with
	linear potential.
	
	Assuming that $|\psi _{0}\rangle $\ is a single-particle eigenstate $|\psi
	_{0}\rangle $ of $H$ with eigenenergy $E_{0}$, $|\psi _{0}\rangle $ can be
	expressed in the form
	
	\begin{equation}
		|\psi _{0}\rangle =\sum_{j=-\infty }^{\infty }f_{j}a_{j}^{\dag }|0\rangle .
	\end{equation}%
	Then other eigenstates can be expressed in the form%
	\begin{equation}
		\left\{ 
		\begin{array}{c}
			|\psi _{l}^{+}\rangle =\sum_{j=-\infty }^{\infty }f_{j-2l}a_{j}^{\dag
			}|0\rangle  \\ 
			|\psi _{l}^{-}\rangle =\sum_{j=-\infty }^{\infty }(-1)^{\text{INT}%
				(j/2)}f_{j-2l}^{\ast }a_{j}^{\dag }|0\rangle 
		\end{array}%
		\right. ,
	\end{equation}%
	with the eigen energy $E_{l}^{\pm }$ in Eq. (\ref{El+-}).\ For an arbitrary
	single-particle initial state%
	\begin{equation}
		|\phi (0)\rangle =\sum_{l}\left( \alpha _{l}|\psi _{l}^{+}\rangle +\beta
		_{l}|\psi _{l}^{-}\rangle \right) ,
	\end{equation}%
	the time evolution of the state at time $t$ is $|\phi (t)\rangle
	=e^{-iHt}|\phi (0)\rangle $. After a sufficient long time, we have%
	\begin{equation}
		|\phi (t)\rangle =e^{-iE_{0}t}\sum_{l}\alpha _{l}e^{-i2l\mathcal{\omega }%
			t}|\psi _{l}^{+}\rangle ,  \label{projected state}
	\end{equation}%
	{which indicates that such a natural time evolution takes the role of
		projection to rule out the component of }$|\psi _{l}^{-}\rangle ${.}
	Obviously, the state $e^{iE_{0}t}|\phi (t)\rangle =\sum_{l}\alpha _{l}e^{-i2l%
		\mathcal{\omega }t}|\psi _{l}^{+}\rangle $\ is a periodic with period $\pi
	/\omega $. We can conclude that, an initial state in the form $\sum_{l}\beta
	_{l}|\psi _{l}^{-}\rangle $ exhibits damping Bloch oscillation, while state $%
	\sum_{l}\alpha _{l}|\psi _{l}^{+}\rangle $\ suffices for growing Bloch
	oscillation. Notably, on the other hand, when we consider a two-particle
	initial state in the form $\sum_{l,l^{\prime }}\gamma _{ll^{\prime }}|\psi
	_{l}^{+}\rangle |\psi _{l^{\prime }}^{-}\rangle $, a standard Bloch
	oscillation occurs. A simple example of such states is $|\phi (t)\rangle
	|\phi (t)\rangle ^{\ast }$\ for large $t$. {We will investigate the
		two-particle dynamics in an alternative way in the next section.} 
	\begin{figure*}[tbh]
		\centering
		\includegraphics[width=0.9\textwidth]{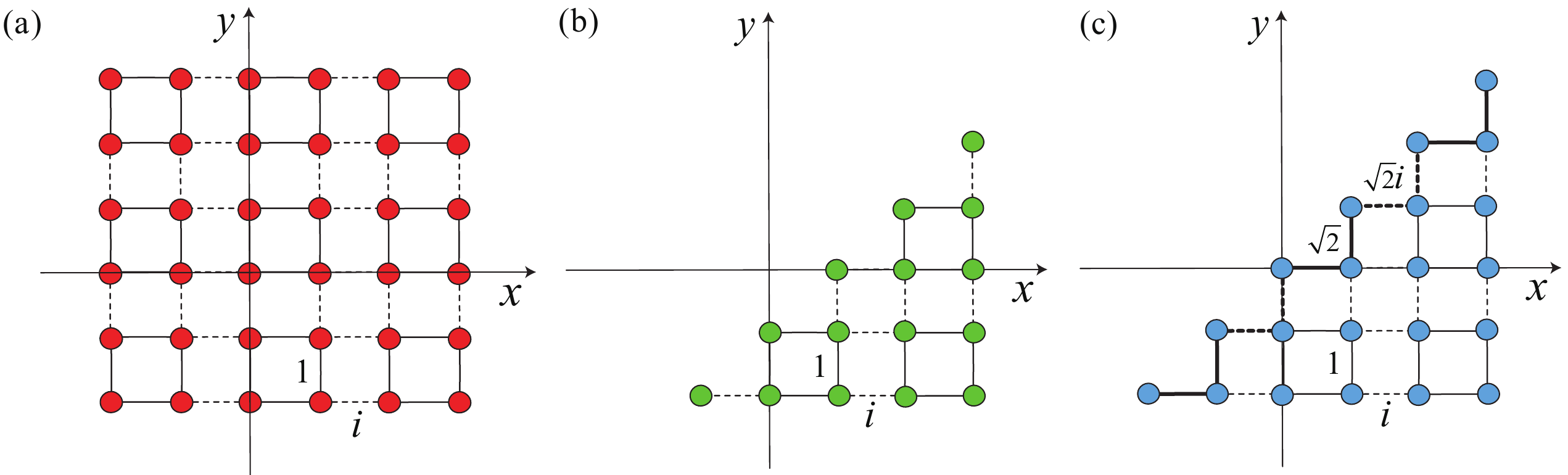}  
		\caption{Schematic illustrations of the Hamiltonians in Eqs.\ (%
			\protect\ref{H2D}), (\protect\ref{H2D+}) and (\protect\ref{H2D-}), which are
			tight-binding models on square lattices with on-site potential $\protect%
			\omega (x+y)$. (a), (b) and (c) can be used to simulate the two-electron,
			two-spinless-fermions, and two-boson Hamiltonian systems shown in Eqs.\ (%
			\protect\ref{He}), (\protect\ref{Hf}) and (\protect\ref{Hb}), respectively.
			It can be observed that (a) has symmetry along the diagonal direction, while
			(b) and (c) can be obtained by the antisymmetric and symmetric
			transformations from (a), respectively.}
		\label{fig4}
	\end{figure*}
	To verify and demonstrate the above analysis, numerical simulations are
	performed to investigate the dynamic behavior driven by the non-Hermitian
	Hamiltonian with a complex Wannier-Stark ladder. We compute the temporal
	evolution for two types of initial states: (i) Gaussian wavepacket state\ in
	the form%
	\begin{equation}
		|\phi (0)\rangle =\sum_{j}e^{-\alpha ^{2}(j-j_{0})^{2}}a_{j}^{\dag
		}|0\rangle ,
	\end{equation}%
	\ $(\alpha =0.3)$; (ii) site state\ at\ $j_{0}$th site in the form $|\phi
	(0)\rangle =a_{j_{0}}^{\dag }|0\rangle $. To present a complete profile of
	the evolved state, we take a mapping $|\phi (t)\rangle \rightarrow $ $%
	e^{-\lambda t}|\phi (t)\rangle $\ by a prefactor to reduce the damping (or
	growing) rate. Obviously, the dynamics exhibit a normal Bloch oscillation
	when taking $\lambda =\func{Im}E_{0}$. For a given initial state, the Dirac
	probability distribution in real space for the evolved state $|\phi
	(t)\rangle $ is 
	\begin{equation}
		P_{n}(t)=\left\vert \left\langle n\right\vert e^{-\lambda t}|\phi (t)\rangle
		\right\vert ^{2}.  \label{Pj(t)}
	\end{equation}%
	We plot $P_{n}(t)$ in Fig.\ \ref{fig3} for several typical parameter values.
	These numerical results agree with our above analysis: (i) when taking $%
	\lambda >\func{Im}E_{0}$, the dynamics are damping periodic; (ii) when
	taking $\lambda =\func{Im}E_{0}$, the dynamics are periodic.
	
	\section{2D simulator for electron-pair dynamics}
	
	\label{2D simulator for electron-pair dynamics} As shown above,
	the non-Hermitian dimerized Hamiltonians always have complex single-particle
	energy levels. The corresponding dynamics are no longer periodic. However,
	some many-particle energy levels\ can be real and equally spaced. In this
	section, we will investigate the Bloch oscillations for a two-particle
	system and propose a 2D simulator to observe such a dynamics in experiments.
	{ We will be considering an electron, spinless fermions, and boson systems,
		respectively. 
		
		(i) For the electron system, the Hamiltonian reads%
		\begin{eqnarray}
			H_{\mathrm{e}} &=&\sum_{j,\sigma =\uparrow ,\downarrow }\left( c_{2j,\sigma
			}^{\dagger }c_{2j+1,\sigma }+\mathrm{H.c.}\right) +i\sum_{j,\sigma =\uparrow
				,\downarrow }\left( c_{2j-1,\sigma }^{\dagger }c_{2j,\sigma }+\mathrm{H.c.}%
			\right)   \notag \\
			&&+\omega \sum_{j,\sigma =\uparrow ,\downarrow }j n_{j,\sigma
			} ,  \label{He}
		\end{eqnarray}%
		where $c_{j,\sigma }$ ($c_{j,\sigma }^{\dagger }$) is the annihilation
		(creation) operator for an electron at site $j$ and $n_{j,\sigma
		}=c_{j,\sigma }^{\dagger }c_{j,\sigma }$. Introducing a local transformation
		on electron operator 
		\begin{eqnarray}
			\mathcal{P}c_{2j+1,\sigma }\mathcal{P}^{-1} &=&(-1)^{j}c_{2j+1,\sigma }, 
			\notag \\
			\mathcal{P}c_{2j,\sigma }\mathcal{P}^{-1} &=&(-1)^{j}c_{2j,\sigma },
		\end{eqnarray}%
		we have 
		\begin{equation}
			\mathcal{PT}H_{\mathrm{e}}\left( \mathcal{PT}\right) ^{-1}=H_{\mathrm{e}},
		\end{equation}%
		i.e., the system has $\mathcal{PT}$\ symmetry.
		
		Now, we focus on the system in a two-electron invariant subsapce with
		opposite spins. Based on the two-electron basis set spanned by $\left\{
		\left\vert x,y\right\rangle \right\} $ where 
		\begin{equation}
			\left\vert x,y\right\rangle =c_{x,\uparrow }^{\dagger }c_{y,\downarrow
			}^{\dagger }\left\vert 0\right\rangle ,
		\end{equation}%
		the Hamiltonian can be expressed as a single-particle Hamiltonian on a
		square lattice \cite{Corrielli2013,Longhi2011}
		
		\begin{eqnarray}
			H_{\mathrm{2D}} &=&\sum_{x,y=-\infty }^{\infty }\left( |2x,y\rangle \langle
			2x+1,y|+\mathrm{H.c.}\right)  \notag \\
			&&+i\sum_{x,y=-\infty }^{\infty }\left( |2x-1,y\rangle \langle 2x,y|+\mathrm{%
				\ \ H.c.}\right)  \notag \\
			&&+\sum_{x,y=-\infty }^{\infty }\left( |x,2y+1\rangle \langle x,2y|+\mathrm{%
				\ H.c.}\right)  \notag \\
			&&+i\sum_{x,y=-\infty }^{\infty }\left( |x,2y-1\rangle \langle x,2y|+\mathrm{%
				\ \ H.c.}\right)  \notag \\
			&&+\omega \sum_{x,y=-\infty }^{\infty }\left( x+y\right) |x,y\rangle \langle
			x,y|.  \label{H2D}
		\end{eqnarray}%
		Fig. \ref{fig4}(a) is a schematics for the structure of $H_{\mathrm{2D}}$.
		One can check that $H_{\mathrm{2D}}$\ has $\mathcal{PT}$\ symmetry, i.e., 
		\begin{equation}
			\left[ \mathcal{PT},H_{\mathrm{2D}}\right] =0,
		\end{equation}%
		due to the fact 
		\begin{equation}
			\mathcal{P}c_{x,\uparrow }^{\dagger }c_{y,\downarrow }^{\dagger }\left\vert
			0\right\rangle =(-1)^{[x/2]+[y/2]}c_{x,\uparrow }^{\dagger }c_{y,\downarrow
			}^{\dagger }\left\vert 0\right\rangle ,
		\end{equation}%
		where $[x]$ represents the greatest integer less than or equal to $x$.
		
		(ii) For the spinless fermion system, the Hamiltonian reads%
		\begin{eqnarray}
			H_{\mathrm{f}} &=&\sum_{j}\left( f_{2j}^{\dagger }f_{2j+1}+\mathrm{H.c.}%
			\right) +i\sum_{j}\left( f_{2j-1}^{\dagger }f_{2j}+\mathrm{H.c.}\right)  
			\notag \\
			&&+\omega \sum_{j}j n_{j} , \label{Hf}
		\end{eqnarray}%
		where $f_{j}$ ($f_{j}^{\dagger }$) is the annihilation (creation) operator
		for an boson at site $j$ and $n_{j}=f_{j}^{\dagger }f_{j}$. The two-boson
		basis set spanned by $\left\{ \left\vert x,y\right\rangle \right\} $ is
		defined as 
		\begin{equation}
			\left\vert x,y\right\rangle =f_{x}^{\dagger }f_{y}^{\dagger }\left\vert
			0\right\rangle ,\,\,(x>y)
		\end{equation}%
		the Hamiltonian can be expressed as a single-particle Hamiltonian on a
		square lattice
		
		\begin{eqnarray}
			H_{\mathrm{2D}}^{+} &=&\sum_{y=-\infty }^{\infty }\sum_{x=[y/2]+1}^{\infty
			}(|2x,y\rangle \langle 2x+1,y|+\mathrm{\ H.c.})  \notag \\
			&&+i\,\sum_{y=-\infty }^{\infty }\sum_{x=[(y+1)/2]+1}^{\infty
			}(|2x-1,y\rangle \langle 2x,y|+\mathrm{\ H.c.})  \notag \\
			&&+\sum_{x=-\infty }^{\infty }\sum_{y=-\infty }^{[x/2]-1}(|x,2y\rangle
			\langle x,2y+1|+\mathrm{\ H.c.})  \notag \\
			&&+i\,\sum_{x=-\infty }^{\infty }\sum_{y=-\infty
			}^{[(x-1)/2]}(|x,2y-1\rangle \langle x,2y|+\mathrm{\ H.c.})  \notag \\
			&&+\omega \sum_{x>y}(x+y)|x,y\rangle \langle x,y|  \label{H2D+}
		\end{eqnarray}%
		Fig. \ref{fig4}(b) is a schematics for the structure of $H_{\mathrm{2D}}^{+}$%
		.
		
		\begin{figure*}[tbh]
			\centering
			\includegraphics[width=0.9\textwidth]{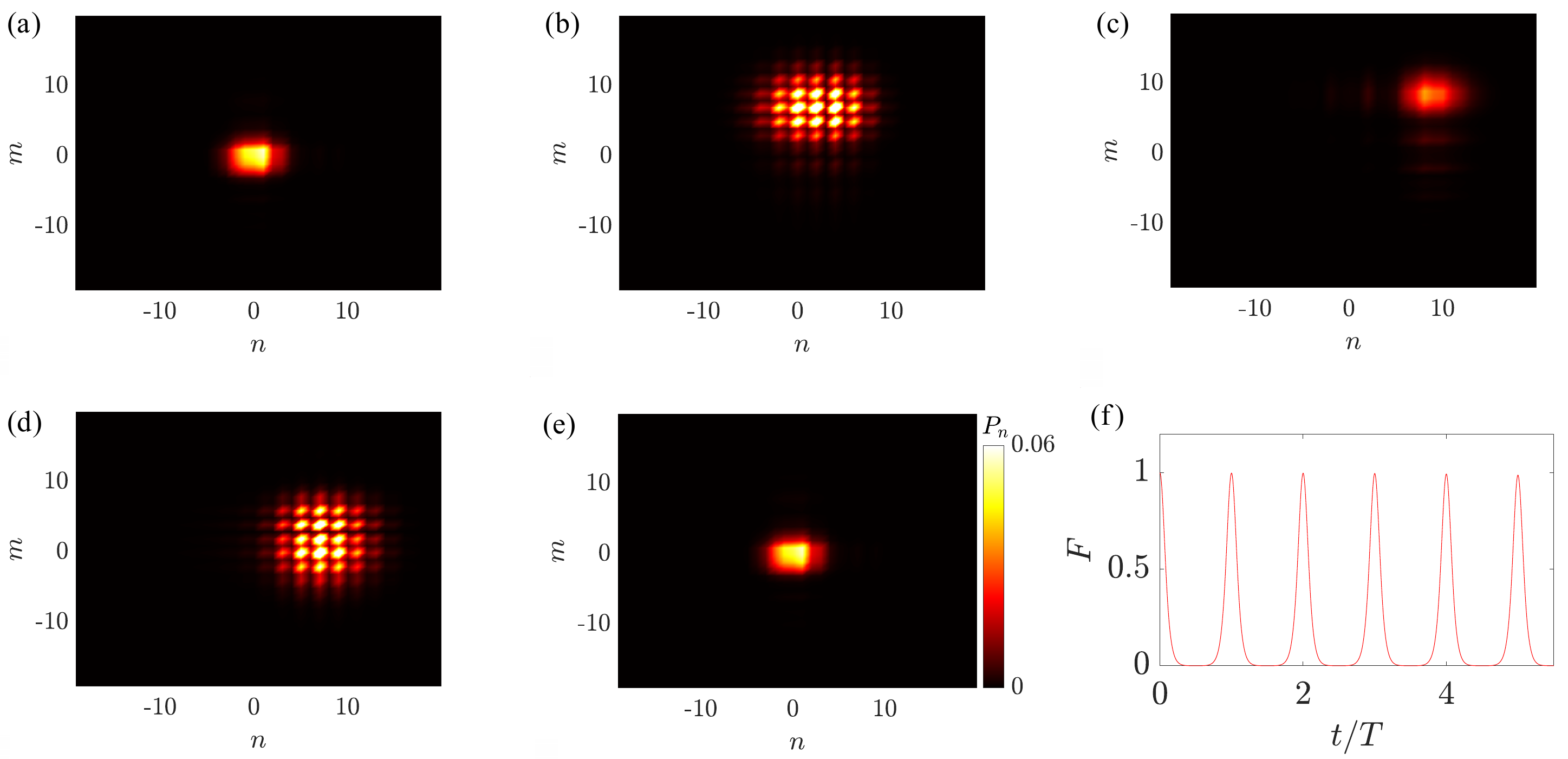}  
			\caption{Plots of $P(x,y,t)$ for several instants defined in Eq.\ (\protect
				\ref{Pxyt}) and $F(t)$\ defined in Eq.\ (\protect\ref{Ft}), obtained by
				numerical diagonalization for the initial state $(-1)^{[y/2]}\protect\mu (x)\protect\mu^* %
				(y)$, where $\protect\mu (x)$ is obtained from the long-time evolved in
				Fig.\ \protect\ref{fig3}(a1). The system parameter $\mathcal{\protect\omega }
				=0.2$.\ Here, (a-e) correspond to the cases with $t=0$, $T/4$, $T/2$, $3T/4$
				and $T $,respectively, and (f) is the plot of fidelity, indicating a
				Bloch oscillation.}
			\label{fig5}
		\end{figure*}
		
		\begin{figure*}[tbh]
			\centering
			\includegraphics[width=0.9\textwidth]{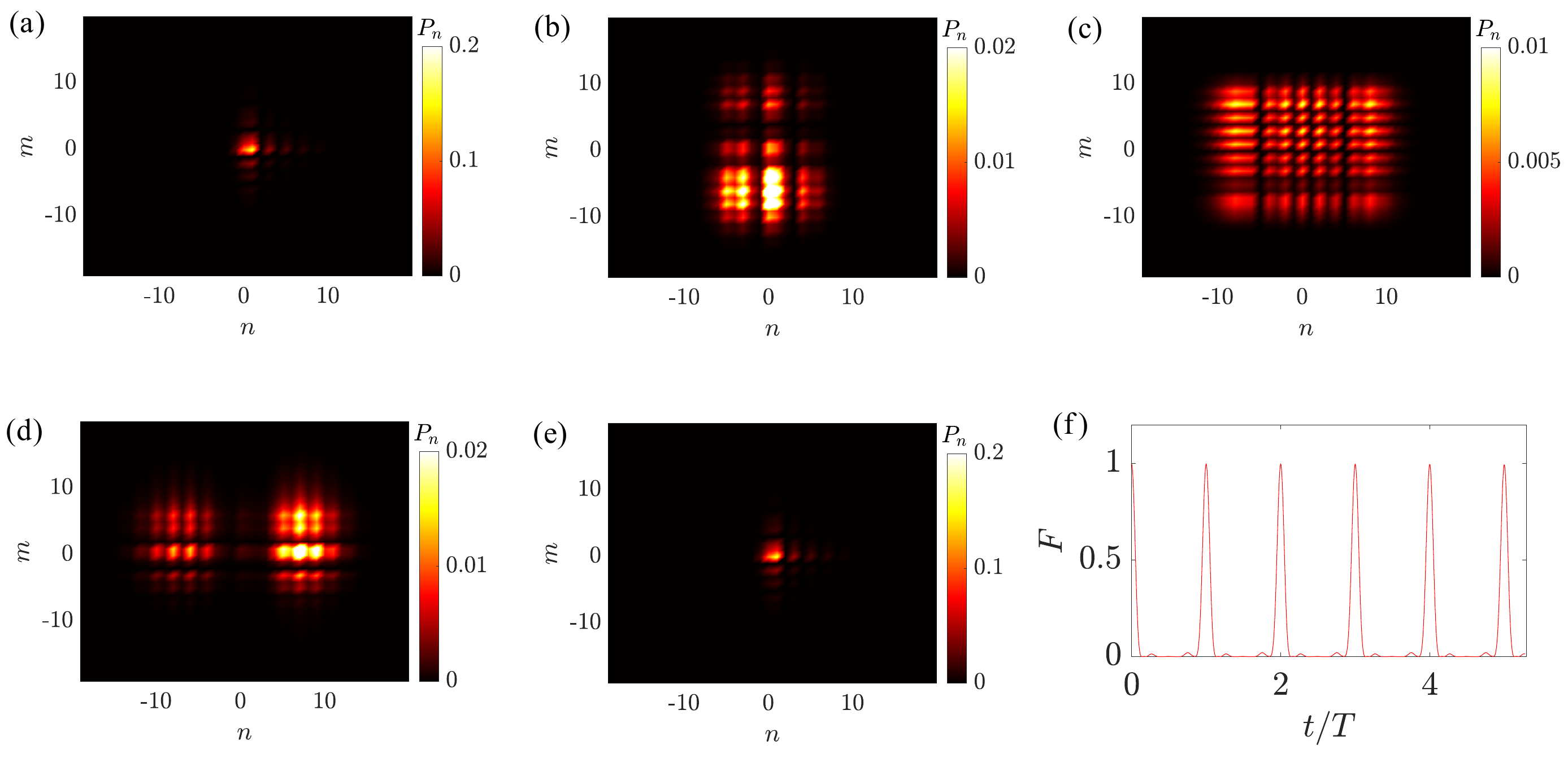}  
			\caption{The same plots as in Fig. \protect\ref{fig5}. Here $\protect\mu (x)$
				is obtained from the result in Fig.\ \protect\ref{fig3}(b1)$.$}
			\label{fig6}
		\end{figure*}
		(iii) For the boson system, the Hamiltonian reads%
		\begin{eqnarray}
			H_{\mathrm{b}} &=&\sum_{j}\left( b_{2j}^{\dagger }b_{2j+1}+\mathrm{H.c.}%
			\right) +i\sum_{j}\left( b_{2j-1}^{\dagger }b_{2j}+\mathrm{H.c.}\right)  
			\notag \\
			&&+\omega \sum_{j}j n_{j} ,  \label{Hb}
		\end{eqnarray}%
		where $b_{j}$ ($b_{j}^{\dagger }$) is the annihilation (creation) operator
		for an boson at site $j$ and $n_{j}=b_{j}^{\dagger }b_{j}$. The two-boson
		basis set spanned by $\left\{ \left\vert x,y\right\rangle \right\} $ is
		defined as 
		\begin{eqnarray}
			\left\vert x,y\right\rangle  &=&b_{x}^{\dagger }b_{y}^{\dagger }\left\vert
			0\right\rangle ,\,\,(x>y)  \notag \\
			|x,y\rangle  &=&\frac{1}{\sqrt{2}}\left( b_{x}^{\dagger }\right) ^{2}\left\vert
			0\right\rangle
			,\,\,(x=y)
		\end{eqnarray}%
		the Hamiltonian can be expressed as a single-particle Hamiltonian on a
		square lattice
		
		\begin{eqnarray}
			H_{\mathrm{2D}}^{-} &=&H_{\mathrm{2D}}^{+}+\sqrt{2}\sum_{y=-\infty }^{\infty
			}(|2y,2y\rangle \langle 2y+1,2y|+\mathrm{\ H.c.})  \notag \\
			&&+\sqrt{2}\sum_{x=-\infty }^{\infty }(|2x+1,2x\rangle \langle 2x+1,2x-1|+ 
			\mathrm{\ H.c.})  \notag \\
			&&+\sqrt{2}i\sum_{y=-\infty }^{\infty }(|2y,2y+1\rangle \langle 2y+1,2y+1|+ 
			\mathrm{\ H.c.})  \notag \\
			&&+\sqrt{2}i\sum_{x=-\infty }^{\infty }(|2x,2x\rangle \langle 2x,2x-1|+ 
			\mathrm{\ H.c.})  \label{H2D-}
		\end{eqnarray}%
		Fig. \ref{fig4}(c) is a schematics for the structure of $H_{\mathrm{2D}}^{-}$%
		.
		
		One can check that the lattice of $H_{\mathrm{2D}}$\ has reflection symmetry
		about the axis $x=y$. Then $H_{\mathrm{2D}}$\ can be decomposed into two
		independent sublattices, symmetric and anti-symmetry. Notably, systems $H_{ 
			\mathrm{2D}}^{+}$\ and $H_{\mathrm{2D}}^{-}$\ are just the two corresponding
		sublattice. The underlying mechanism is the following relation. The singlet
		two-electron states 
		\begin{equation}
			\left\{ 
			\begin{array}{cc}
				\frac{1}{\sqrt{2}}\left( c_{x,\uparrow }^{\dagger }c_{y,\downarrow
				}^{\dagger }-c_{x,\downarrow }^{\dagger }c_{y,\uparrow }^{\dagger }\right)
				\left\vert 0\right\rangle , & x\neq y \\ 
				c_{x,\uparrow }^{\dagger }c_{x,\downarrow }^{\dagger }\left\vert
				0\right\rangle , & x=y%
			\end{array}
			\right. ,
		\end{equation}%
		for $H_{\mathrm{e}}$\ are equivalent to the two-boson states%
		\begin{equation}
			\left\vert x,y\right\rangle =\left\{ 
			\begin{array}{cc}
				b_{x}^{\dagger }b_{y}^{\dagger }\left\vert 0\right\rangle & x\neq y \\ 
				\frac{1}{\sqrt{2}}\left( b_{x}^{\dagger }\right) ^{2}\left\vert
				0\right\rangle , & x=y%
			\end{array}
			\right. ,
		\end{equation}%
		for $H_{\mathrm{b}}$. In parallel, the triplet two-electron states 
		\begin{equation}
			\left\vert x,y\right\rangle =\frac{1}{\sqrt{2}}\left( c_{x,\uparrow
			}^{\dagger }c_{y,\downarrow }^{\dagger }+c_{x,\downarrow }^{\dagger
			}c_{y,\uparrow }^{\dagger }\right) \left\vert 0\right\rangle ,
		\end{equation}%
		for $H_{\mathrm{e}}$ are equivalent to the two-spinless-fermion states%
		\begin{equation}
			\left\vert x,y\right\rangle =c_{x}^{\dagger }c_{y}^{\dagger }\left\vert
			0\right\rangle ,
		\end{equation}%
		for $H_{\mathrm{f}}$. In this sense, the physical properties of both
		two-boson and two-spinless-fermion systems can be derived from those of the $%
		H_{\mathrm{2D}}$ system.}
	
	In experiments, single-particle hopping dynamics can be simulated by
	discretized spatial light transport in an engineered 2D square lattice of
	evanescently coupled optical waveguides \cite{Christodoulides2003}. A 2D
	lattice can be fabricated by coupled waveguides, by which the temporal
	evolution of the single-particle probability distribution\ in the 2D lattice
	can be visualized by the spatial propagation of the light intensity.
	According to the non-Hermitian quantum theory, an eigenstate with real
	energy can be written as the form of $\mathcal{PT}$\ symmetry.
	
	We compute the temporal evolution for 2D initial states in the product form 
	\begin{equation}
		|\Phi (0)\rangle =|\varphi _{x}\rangle |\varphi _{y}\rangle ,
	\end{equation}%
	where $|\varphi _{x}\rangle $ and $|\varphi _{y}\rangle $ are two normalized
	states for 1D,%
	\begin{eqnarray}
		|\varphi _{x}\rangle &=&\sum_{x}\mu (x)|x\rangle ,\, \\
		|\varphi _{y}\rangle &=&\mathcal{T}g\sum_{y}\mu (y)|y\rangle
		=\sum_{y}(-1)^{[y/2]}\mu ^{\ast }(y)|y\rangle .
	\end{eqnarray}%
	Here, the coefficients $\mu (x)$\ can be extracted from\ the time evolution
	for 1D system\ with initial states $|\phi (0)\rangle $\ based on the
	mechanism in Eq. (\ref{projected state}). We still employ two types of
	states, the Gaussian wavepacket state\ and the site state\ at\ $j_{0}$th
	site.
	
	{We calculate the Dirac probability distribution in real space} 
	\begin{equation}
		P(x,y,t)=\left\vert \left\langle x,y\right\vert e^{-iH_{\mathrm{2D}}t}|\Phi
		(0)\rangle \right\vert ^{2},  \label{Pxyt}
	\end{equation}%
	and the fidelity
	
	\begin{equation}
		F(t)=\frac{\left\vert \left\langle \Phi (0)\right\vert e^{-iH_{\mathrm{2D}
				}t}|\Phi (0)\rangle \right\vert ^{2}}{\left\vert e^{-iH_{\mathrm{2D}}t}|\Phi
			(0)\rangle \right\vert ^{2}\left\vert |\Phi (0)\rangle \right\vert ^{2}},
		\label{Ft}
	\end{equation}%
	for time evolution. We plot $P(x,y,t)$ and $F(t) $\ in Fig. \ref{fig5} and %
	\ref{fig6}\ for several typical parameter values. {These  numerical results
		agree with our above analysis that the dynamics are periodic for
		proper initial states. Therefore, we conclude that our result  for extended
		Wannier-Stark ladder in 1D system can also be applied to a 2D  system.}
	
	\section{Summary}
	
	\label{Summary}
	
	In summary, our research has investigated the impact of a linear potential
	on arbitrary periodic systems within the tight-binding model framework,
	irrespective of the Hermiticity of the system. We have extended the concept
	of the Wannier-Stark ladder to the complex domain, enabling the analytical
	prediction of the dynamics for numerous non-Hermitian systems with a
	constant particle number. We propose two types of dimerized non-Hermitian
	systems to substantiate our findings. Both analytical analysis and numerical
	simulations indicate that these dimerized non-Hermitian systems support
	single-particle Bloch oscillations characterized by a damping (or growing)
	rate, as well as standard two-particle Bloch oscillations under specific
	conditions. Furthermore, we demonstrate that the concept of the extended
	Wannier-Stark ladder, initially developed for 1D systems, is also applicable
	to 2D systems. {We introduce a scheme to illustrate these results
		through particle-pair dynamics} within a single-particle 2D $\mathcal{PT}$%
	-symmetric square lattice. These findings lay the groundwork for exploring a
	diverse range of periodic systems subjected to a linear potential. The
	ability to predict the dynamics of such systems analytically is a valuable
	advancement in the field of condensed matter physics and quantum mechanics.
	
	\acknowledgments This work was supported by the National Natural Science
	Foundation of China (under Grant No. 12374461).

   \bibliography{References}

\begin{thebibliography}{38}%
\makeatletter
\providecommand \@ifxundefined [1]{%
 \@ifx{#1\undefined}
}%
\providecommand \@ifnum [1]{%
 \ifnum #1\expandafter \@firstoftwo
 \else \expandafter \@secondoftwo
 \fi
}%
\providecommand \@ifx [1]{%
 \ifx #1\expandafter \@firstoftwo
 \else \expandafter \@secondoftwo
 \fi
}%
\providecommand \natexlab [1]{#1}%
\providecommand \enquote  [1]{``#1''}%
\providecommand \bibnamefont  [1]{#1}%
\providecommand \bibfnamefont [1]{#1}%
\providecommand \citenamefont [1]{#1}%
\providecommand \href@noop [0]{\@secondoftwo}%
\providecommand \href [0]{\begingroup \@sanitize@url \@href}%
\providecommand \@href[1]{\@@startlink{#1}\@@href}%
\providecommand \@@href[1]{\endgroup#1\@@endlink}%
\providecommand \@sanitize@url [0]{\catcode `\\12\catcode `\$12\catcode
  `\&12\catcode `\#12\catcode `\^12\catcode `\_12\catcode `\%12\relax}%
\providecommand \@@startlink[1]{}%
\providecommand \@@endlink[0]{}%
\providecommand \url  [0]{\begingroup\@sanitize@url \@url }%
\providecommand \@url [1]{\endgroup\@href {#1}{\urlprefix }}%
\providecommand \urlprefix  [0]{URL }%
\providecommand \Eprint [0]{\href }%
\providecommand \doibase [0]{http://dx.doi.org/}%
\providecommand \selectlanguage [0]{\@gobble}%
\providecommand \bibinfo  [0]{\@secondoftwo}%
\providecommand \bibfield  [0]{\@secondoftwo}%
\providecommand \translation [1]{[#1]}%
\providecommand \BibitemOpen [0]{}%
\providecommand \bibitemStop [0]{}%
\providecommand \bibitemNoStop [0]{.\EOS\space}%
\providecommand \EOS [0]{\spacefactor3000\relax}%
\providecommand \BibitemShut  [1]{\csname bibitem#1\endcsname}%
\let\auto@bib@innerbib\@empty
\bibitem [{\citenamefont {Bloch}(1929)}]{Bloch1929}%
  \BibitemOpen
  \bibfield  {author} {\bibinfo {author} {\bibfnamefont {Felix}\ \bibnamefont
  {Bloch}},\ }\bibfield  {title} {\enquote {\bibinfo {title} {{\"Uber die
  Quantenmechanik der Elektronen in Kristallgittern}},}\ }\href {\doibase
  10.1007/bf01339455} {\bibfield  {journal} {\bibinfo  {journal} {Zeitschrift
  f\"ur Physik}\ }\textbf {\bibinfo {volume} {52}},\ \bibinfo {pages}
  {555--600} (\bibinfo {year} {1929})}\BibitemShut {NoStop}%
\bibitem [{\citenamefont {Wannier}(1959)}]{wannier1959elements}%
  \BibitemOpen
  \bibfield  {author} {\bibinfo {author} {\bibfnamefont {Gregory~H}\
  \bibnamefont {Wannier}},\ }\href@noop {} {\emph {\bibinfo {title} {{Elements
  of solid state theory}}}}\ (\bibinfo  {publisher} {CUP Archive},\ \bibinfo
  {year} {1959})\BibitemShut {NoStop}%
\bibitem [{\citenamefont {Wannier}(1960)}]{Wannier1960}%
  \BibitemOpen
  \bibfield  {author} {\bibinfo {author} {\bibfnamefont {Gregory~H.}\
  \bibnamefont {Wannier}},\ }\bibfield  {title} {\enquote {\bibinfo {title}
  {{Wave Functions and Effective Hamiltonian for Bloch Electrons in an Electric
  Field}},}\ }\href {\doibase 10.1103/physrev.117.432} {\bibfield  {journal}
  {\bibinfo  {journal} {Physical Review}\ }\textbf {\bibinfo {volume} {117}},\
  \bibinfo {pages} {432--439} (\bibinfo {year} {1960})}\BibitemShut {NoStop}%
\bibitem [{\citenamefont {Glück}(2002)}]{Glueck2002}%
  \BibitemOpen
  \bibfield  {author} {\bibinfo {author} {\bibfnamefont {M}~\bibnamefont
  {Glück}},\ }\bibfield  {title} {\enquote {\bibinfo {title} {{Wannier–Stark
  resonances in optical and semiconductor superlattices}},}\ }\href {\doibase
  10.1016/s0370-1573(02)00142-4} {\bibfield  {journal} {\bibinfo  {journal}
  {Physics Reports}\ }\textbf {\bibinfo {volume} {366}},\ \bibinfo {pages}
  {103--182} (\bibinfo {year} {2002})}\BibitemShut {NoStop}%
\bibitem [{\citenamefont {Waschke}\ \emph {et~al.}(1993)\citenamefont
  {Waschke}, \citenamefont {Roskos}, \citenamefont {Schwedler}, \citenamefont
  {Leo}, \citenamefont {Kurz},\ and\ \citenamefont {Köhler}}]{Waschke1993}%
  \BibitemOpen
  \bibfield  {author} {\bibinfo {author} {\bibfnamefont {Christian}\
  \bibnamefont {Waschke}}, \bibinfo {author} {\bibfnamefont {Hartmut~G.}\
  \bibnamefont {Roskos}}, \bibinfo {author} {\bibfnamefont {Ralf}\ \bibnamefont
  {Schwedler}}, \bibinfo {author} {\bibfnamefont {Karl}\ \bibnamefont {Leo}},
  \bibinfo {author} {\bibfnamefont {Heinrich}\ \bibnamefont {Kurz}}, \ and\
  \bibinfo {author} {\bibfnamefont {Klaus}\ \bibnamefont {Köhler}},\
  }\bibfield  {title} {\enquote {\bibinfo {title} {{Coherent submillimeter-wave
  emission from Bloch oscillations in a semiconductor superlattice}},}\ }\href
  {\doibase 10.1103/physrevlett.70.3319} {\bibfield  {journal} {\bibinfo
  {journal} {Physical Review Letters}\ }\textbf {\bibinfo {volume} {70}},\
  \bibinfo {pages} {3319--3322} (\bibinfo {year} {1993})}\BibitemShut {NoStop}%
\bibitem [{\citenamefont {Ben~Dahan}\ \emph {et~al.}(1996)\citenamefont
  {Ben~Dahan}, \citenamefont {Peik}, \citenamefont {Reichel}, \citenamefont
  {Castin},\ and\ \citenamefont {Salomon}}]{BenDahan1996}%
  \BibitemOpen
  \bibfield  {author} {\bibinfo {author} {\bibfnamefont {Maxime}\ \bibnamefont
  {Ben~Dahan}}, \bibinfo {author} {\bibfnamefont {Ekkehard}\ \bibnamefont
  {Peik}}, \bibinfo {author} {\bibfnamefont {Jakob}\ \bibnamefont {Reichel}},
  \bibinfo {author} {\bibfnamefont {Yvan}\ \bibnamefont {Castin}}, \ and\
  \bibinfo {author} {\bibfnamefont {Christophe}\ \bibnamefont {Salomon}},\
  }\bibfield  {title} {\enquote {\bibinfo {title} {{Bloch Oscillations of Atoms
  in an Optical Potential}},}\ }\href {\doibase 10.1103/physrevlett.76.4508}
  {\bibfield  {journal} {\bibinfo  {journal} {Physical Review Letters}\
  }\textbf {\bibinfo {volume} {76}},\ \bibinfo {pages} {4508--4511} (\bibinfo
  {year} {1996})}\BibitemShut {NoStop}%
\bibitem [{\citenamefont {Wilkinson}\ \emph {et~al.}(1996)\citenamefont
  {Wilkinson}, \citenamefont {Bharucha}, \citenamefont {Madison}, \citenamefont
  {Niu},\ and\ \citenamefont {Raizen}}]{Wilkinson1996}%
  \BibitemOpen
  \bibfield  {author} {\bibinfo {author} {\bibfnamefont {S.~R.}\ \bibnamefont
  {Wilkinson}}, \bibinfo {author} {\bibfnamefont {C.~F.}\ \bibnamefont
  {Bharucha}}, \bibinfo {author} {\bibfnamefont {K.~W.}\ \bibnamefont
  {Madison}}, \bibinfo {author} {\bibfnamefont {Qian}\ \bibnamefont {Niu}}, \
  and\ \bibinfo {author} {\bibfnamefont {M.~G.}\ \bibnamefont {Raizen}},\
  }\bibfield  {title} {\enquote {\bibinfo {title} {{Observation of Atomic
  Wannier-Stark Ladders in an Accelerating Optical Potential}},}\ }\href
  {\doibase 10.1103/physrevlett.76.4512} {\bibfield  {journal} {\bibinfo
  {journal} {Physical Review Letters}\ }\textbf {\bibinfo {volume} {76}},\
  \bibinfo {pages} {4512--4515} (\bibinfo {year} {1996})}\BibitemShut {NoStop}%
\bibitem [{\citenamefont {Anderson}\ and\ \citenamefont
  {Kasevich}(1998)}]{Anderson1998}%
  \BibitemOpen
  \bibfield  {author} {\bibinfo {author} {\bibfnamefont {B.~P.}\ \bibnamefont
  {Anderson}}\ and\ \bibinfo {author} {\bibfnamefont {M.~A.}\ \bibnamefont
  {Kasevich}},\ }\bibfield  {title} {\enquote {\bibinfo {title} {{Macroscopic
  Quantum Interference from Atomic Tunnel Arrays}},}\ }\href {\doibase
  10.1126/science.282.5394.1686} {\bibfield  {journal} {\bibinfo  {journal}
  {Science}\ }\textbf {\bibinfo {volume} {282}},\ \bibinfo {pages} {1686--1689}
  (\bibinfo {year} {1998})}\BibitemShut {NoStop}%
\bibitem [{\citenamefont {Morsch}\ \emph {et~al.}(2001)\citenamefont {Morsch},
  \citenamefont {Müller}, \citenamefont {Cristiani}, \citenamefont
  {Ciampini},\ and\ \citenamefont {Arimondo}}]{Morsch2001}%
  \BibitemOpen
  \bibfield  {author} {\bibinfo {author} {\bibfnamefont {O.}~\bibnamefont
  {Morsch}}, \bibinfo {author} {\bibfnamefont {J.~H.}\ \bibnamefont {Müller}},
  \bibinfo {author} {\bibfnamefont {M.}~\bibnamefont {Cristiani}}, \bibinfo
  {author} {\bibfnamefont {D.}~\bibnamefont {Ciampini}}, \ and\ \bibinfo
  {author} {\bibfnamefont {E.}~\bibnamefont {Arimondo}},\ }\bibfield  {title}
  {\enquote {\bibinfo {title} {{Bloch Oscillations and Mean-Field Effects of
  Bose-Einstein Condensates in 1D Optical Lattices}},}\ }\href {\doibase
  10.1103/physrevlett.87.140402} {\bibfield  {journal} {\bibinfo  {journal}
  {Physical Review Letters}\ }\textbf {\bibinfo {volume} {87}},\ \bibinfo
  {pages} {140402} (\bibinfo {year} {2001})}\BibitemShut {NoStop}%
\bibitem [{\citenamefont {Morandotti}\ \emph {et~al.}(1999)\citenamefont
  {Morandotti}, \citenamefont {Peschel}, \citenamefont {Aitchison},
  \citenamefont {Eisenberg},\ and\ \citenamefont
  {Silberberg}}]{Morandotti1999}%
  \BibitemOpen
  \bibfield  {author} {\bibinfo {author} {\bibfnamefont {R.}~\bibnamefont
  {Morandotti}}, \bibinfo {author} {\bibfnamefont {U.}~\bibnamefont {Peschel}},
  \bibinfo {author} {\bibfnamefont {J.~S.}\ \bibnamefont {Aitchison}}, \bibinfo
  {author} {\bibfnamefont {H.~S.}\ \bibnamefont {Eisenberg}}, \ and\ \bibinfo
  {author} {\bibfnamefont {Y.}~\bibnamefont {Silberberg}},\ }\bibfield  {title}
  {\enquote {\bibinfo {title} {{Experimental Observation of Linear and
  Nonlinear Optical Bloch Oscillations}},}\ }\href {\doibase
  10.1103/physrevlett.83.4756} {\bibfield  {journal} {\bibinfo  {journal}
  {Physical Review Letters}\ }\textbf {\bibinfo {volume} {83}},\ \bibinfo
  {pages} {4756--4759} (\bibinfo {year} {1999})}\BibitemShut {NoStop}%
\bibitem [{\citenamefont {Sanchis-Alepuz}\ \emph {et~al.}(2007)\citenamefont
  {Sanchis-Alepuz}, \citenamefont {Kosevich},\ and\ \citenamefont
  {Sánchez-Dehesa}}]{SanchisAlepuz2007}%
  \BibitemOpen
  \bibfield  {author} {\bibinfo {author} {\bibfnamefont {Helios}\ \bibnamefont
  {Sanchis-Alepuz}}, \bibinfo {author} {\bibfnamefont {Yuriy~A.}\ \bibnamefont
  {Kosevich}}, \ and\ \bibinfo {author} {\bibfnamefont {José}\ \bibnamefont
  {Sánchez-Dehesa}},\ }\bibfield  {title} {\enquote {\bibinfo {title}
  {{Acoustic Analogue of Electronic Bloch Oscillations and Resonant Zener
  Tunneling in Ultrasonic Superlattices}},}\ }\href {\doibase
  10.1103/physrevlett.98.134301} {\bibfield  {journal} {\bibinfo  {journal}
  {Physical Review Letters}\ }\textbf {\bibinfo {volume} {98}},\ \bibinfo
  {pages} {134301} (\bibinfo {year} {2007})}\BibitemShut {NoStop}%
\bibitem [{\citenamefont {Meinert}\ \emph {et~al.}(2017)\citenamefont
  {Meinert}, \citenamefont {Knap}, \citenamefont {Kirilov}, \citenamefont
  {Jag-Lauber}, \citenamefont {Zvonarev}, \citenamefont {Demler},\ and\
  \citenamefont {Nägerl}}]{Meinert2017}%
  \BibitemOpen
  \bibfield  {author} {\bibinfo {author} {\bibfnamefont {Florian}\ \bibnamefont
  {Meinert}}, \bibinfo {author} {\bibfnamefont {Michael}\ \bibnamefont {Knap}},
  \bibinfo {author} {\bibfnamefont {Emil}\ \bibnamefont {Kirilov}}, \bibinfo
  {author} {\bibfnamefont {Katharina}\ \bibnamefont {Jag-Lauber}}, \bibinfo
  {author} {\bibfnamefont {Mikhail~B.}\ \bibnamefont {Zvonarev}}, \bibinfo
  {author} {\bibfnamefont {Eugene}\ \bibnamefont {Demler}}, \ and\ \bibinfo
  {author} {\bibfnamefont {Hanns-Christoph}\ \bibnamefont {Nägerl}},\
  }\bibfield  {title} {\enquote {\bibinfo {title} {{Bloch oscillations in the
  absence of a lattice}},}\ }\href {\doibase 10.1126/science.aah6616}
  {\bibfield  {journal} {\bibinfo  {journal} {Science}\ }\textbf {\bibinfo
  {volume} {356}},\ \bibinfo {pages} {945--948} (\bibinfo {year}
  {2017})}\BibitemShut {NoStop}%
\bibitem [{\citenamefont {Zhang}\ \emph {et~al.}(2022)\citenamefont {Zhang},
  \citenamefont {Yuan}, \citenamefont {Wang}, \citenamefont {Di}, \citenamefont
  {Sun}, \citenamefont {Zheng}, \citenamefont {Sun},\ and\ \citenamefont
  {Zhang}}]{Zhang2022}%
  \BibitemOpen
  \bibfield  {author} {\bibinfo {author} {\bibfnamefont {Weixuan}\ \bibnamefont
  {Zhang}}, \bibinfo {author} {\bibfnamefont {Hao}\ \bibnamefont {Yuan}},
  \bibinfo {author} {\bibfnamefont {Haiteng}\ \bibnamefont {Wang}}, \bibinfo
  {author} {\bibfnamefont {Fengxiao}\ \bibnamefont {Di}}, \bibinfo {author}
  {\bibfnamefont {Na}~\bibnamefont {Sun}}, \bibinfo {author} {\bibfnamefont
  {Xingen}\ \bibnamefont {Zheng}}, \bibinfo {author} {\bibfnamefont {Houjun}\
  \bibnamefont {Sun}}, \ and\ \bibinfo {author} {\bibfnamefont {Xiangdong}\
  \bibnamefont {Zhang}},\ }\bibfield  {title} {\enquote {\bibinfo {title}
  {{Observation of Bloch oscillations dominated by effective anyonic particle
  statistics}},}\ }\href {\doibase 10.1038/s41467-022-29895-0} {\bibfield
  {journal} {\bibinfo  {journal} {Nature Communications}\ }\textbf {\bibinfo
  {volume} {13}} (\bibinfo {year} {2022}),\
  10.1038/s41467-022-29895-0}\BibitemShut {NoStop}%
\bibitem [{\citenamefont {Hansen}\ \emph {et~al.}(2022)\citenamefont {Hansen},
  \citenamefont {Syljuåsen}, \citenamefont {Jensen}, \citenamefont
  {Schäffer}, \citenamefont {Andersen}, \citenamefont {Boehm}, \citenamefont
  {Rodriguez-Rivera}, \citenamefont {Christensen},\ and\ \citenamefont
  {Lefmann}}]{Hansen2022}%
  \BibitemOpen
  \bibfield  {author} {\bibinfo {author} {\bibfnamefont {Ursula~B.}\
  \bibnamefont {Hansen}}, \bibinfo {author} {\bibfnamefont {Olav~F.}\
  \bibnamefont {Syljuåsen}}, \bibinfo {author} {\bibfnamefont {Jens}\
  \bibnamefont {Jensen}}, \bibinfo {author} {\bibfnamefont {Turi~K.}\
  \bibnamefont {Schäffer}}, \bibinfo {author} {\bibfnamefont {Christopher~R.}\
  \bibnamefont {Andersen}}, \bibinfo {author} {\bibfnamefont {Martin}\
  \bibnamefont {Boehm}}, \bibinfo {author} {\bibfnamefont {Jose~A.}\
  \bibnamefont {Rodriguez-Rivera}}, \bibinfo {author} {\bibfnamefont
  {Niels~B.}\ \bibnamefont {Christensen}}, \ and\ \bibinfo {author}
  {\bibfnamefont {Kim}\ \bibnamefont {Lefmann}},\ }\bibfield  {title} {\enquote
  {\bibinfo {title} {{Magnetic Bloch oscillations and domain wall dynamics in a
  near-Ising ferromagnetic chain}},}\ }\href {\doibase
  10.1038/s41467-022-29854-9} {\bibfield  {journal} {\bibinfo  {journal}
  {Nature Communications}\ }\textbf {\bibinfo {volume} {13}} (\bibinfo {year}
  {2022}),\ 10.1038/s41467-022-29854-9}\BibitemShut {NoStop}%
\bibitem [{\citenamefont {Longhi}(2009)}]{Longhi2009}%
  \BibitemOpen
  \bibfield  {author} {\bibinfo {author} {\bibfnamefont {S.}~\bibnamefont
  {Longhi}},\ }\bibfield  {title} {\enquote {\bibinfo {title} {{Bloch
  Oscillations in Complex Crystals with $\mathcal{PT}$ Symmetry}},}\ }\href
  {\doibase 10.1103/physrevlett.103.123601} {\bibfield  {journal} {\bibinfo
  {journal} {Physical Review Letters}\ }\textbf {\bibinfo {volume} {103}},\
  \bibinfo {pages} {123601} (\bibinfo {year} {2009})}\BibitemShut {NoStop}%
\bibitem [{\citenamefont {Longhi}(2014)}]{Longhi2014}%
  \BibitemOpen
  \bibfield  {author} {\bibinfo {author} {\bibfnamefont {S.}~\bibnamefont
  {Longhi}},\ }\bibfield  {title} {\enquote {\bibinfo {title} {{Exceptional
  points and Bloch oscillations in non-Hermitian lattices with unidirectional
  hopping}},}\ }\href {\doibase 10.1209/0295-5075/106/34001} {\bibfield
  {journal} {\bibinfo  {journal} {EPL (Europhysics Letters)}\ }\textbf
  {\bibinfo {volume} {106}},\ \bibinfo {pages} {34001} (\bibinfo {year}
  {2014})}\BibitemShut {NoStop}%
\bibitem [{\citenamefont {Longhi}(2015)}]{Longhi2015}%
  \BibitemOpen
  \bibfield  {author} {\bibinfo {author} {\bibfnamefont {Stefano}\ \bibnamefont
  {Longhi}},\ }\bibfield  {title} {\enquote {\bibinfo {title} {{Bloch
  oscillations in non-Hermitian lattices with trajectories in the complex
  plane}},}\ }\href {\doibase 10.1103/physreva.92.042116} {\bibfield  {journal}
  {\bibinfo  {journal} {Physical Review A}\ }\textbf {\bibinfo {volume} {92}},\
  \bibinfo {pages} {042116} (\bibinfo {year} {2015})}\BibitemShut {NoStop}%
\bibitem [{\citenamefont {Graefe}\ \emph {et~al.}(2016)\citenamefont {Graefe},
  \citenamefont {Korsch},\ and\ \citenamefont {Rush}}]{Graefe2016}%
  \BibitemOpen
  \bibfield  {author} {\bibinfo {author} {\bibfnamefont {E~M}\ \bibnamefont
  {Graefe}}, \bibinfo {author} {\bibfnamefont {H~J}\ \bibnamefont {Korsch}}, \
  and\ \bibinfo {author} {\bibfnamefont {A}~\bibnamefont {Rush}},\ }\bibfield
  {title} {\enquote {\bibinfo {title} {{Quasiclassical analysis of Bloch
  oscillations in non-Hermitian tight-binding lattices}},}\ }\href {\doibase
  10.1088/1367-2630/18/7/075009} {\bibfield  {journal} {\bibinfo  {journal}
  {New Journal of Physics}\ }\textbf {\bibinfo {volume} {18}},\ \bibinfo
  {pages} {075009} (\bibinfo {year} {2016})}\BibitemShut {NoStop}%
\bibitem [{\citenamefont {Longhi}(2017)}]{Longhi2017}%
  \BibitemOpen
  \bibfield  {author} {\bibinfo {author} {\bibfnamefont {Stefano}\ \bibnamefont
  {Longhi}},\ }\bibfield  {title} {\enquote {\bibinfo {title} {{Non-Hermitian
  bidirectional robust transport}},}\ }\href {\doibase
  10.1103/physrevb.95.014201} {\bibfield  {journal} {\bibinfo  {journal}
  {Physical Review B}\ }\textbf {\bibinfo {volume} {95}},\ \bibinfo {pages}
  {014201} (\bibinfo {year} {2017})}\BibitemShut {NoStop}%
\bibitem [{\citenamefont {Lee}\ and\ \citenamefont {Chan}(2014)}]{Lee2014}%
  \BibitemOpen
  \bibfield  {author} {\bibinfo {author} {\bibfnamefont {Tony~E.}\ \bibnamefont
  {Lee}}\ and\ \bibinfo {author} {\bibfnamefont {Ching-Kit}\ \bibnamefont
  {Chan}},\ }\bibfield  {title} {\enquote {\bibinfo {title} {{Heralded
  Magnetism in Non-Hermitian Atomic Systems}},}\ }\href {\doibase
  10.1103/physrevx.4.041001} {\bibfield  {journal} {\bibinfo  {journal}
  {Physical Review X}\ }\textbf {\bibinfo {volume} {4}},\ \bibinfo {pages}
  {041001} (\bibinfo {year} {2014})}\BibitemShut {NoStop}%
\bibitem [{\citenamefont {Bender}\ and\ \citenamefont
  {Boettcher}(1998)}]{Bender1998}%
  \BibitemOpen
  \bibfield  {author} {\bibinfo {author} {\bibfnamefont {Carl~M.}\ \bibnamefont
  {Bender}}\ and\ \bibinfo {author} {\bibfnamefont {Stefan}\ \bibnamefont
  {Boettcher}},\ }\bibfield  {title} {\enquote {\bibinfo {title} {{Real Spectra
  in Non-Hermitian Hamiltonians Having $\mathcal{PT}$ Symmetry}},}\ }\href
  {\doibase 10.1103/physrevlett.80.5243} {\bibfield  {journal} {\bibinfo
  {journal} {Physical Review Letters}\ }\textbf {\bibinfo {volume} {80}},\
  \bibinfo {pages} {5243--5246} (\bibinfo {year} {1998})}\BibitemShut {NoStop}%
\bibitem [{\citenamefont {Mostafazadeh}(2003)}]{Mostafazadeh2003}%
  \BibitemOpen
  \bibfield  {author} {\bibinfo {author} {\bibfnamefont {Ali}\ \bibnamefont
  {Mostafazadeh}},\ }\bibfield  {title} {\enquote {\bibinfo {title}
  {{ExactPT-symmetry is equivalent to Hermiticity}},}\ }\href {\doibase
  10.1088/0305-4470/36/25/312} {\bibfield  {journal} {\bibinfo  {journal}
  {Journal of Physics A: Mathematical and General}\ }\textbf {\bibinfo {volume}
  {36}},\ \bibinfo {pages} {7081--7091} (\bibinfo {year} {2003})}\BibitemShut
  {NoStop}%
\bibitem [{\citenamefont {Mostafazadeh}\ and\ \citenamefont
  {Batal}(2004)}]{Mostafazadeh2004}%
  \BibitemOpen
  \bibfield  {author} {\bibinfo {author} {\bibfnamefont {Ali}\ \bibnamefont
  {Mostafazadeh}}\ and\ \bibinfo {author} {\bibfnamefont {Ahmet}\ \bibnamefont
  {Batal}},\ }\bibfield  {title} {\enquote {\bibinfo {title} {{Physical aspects
  of pseudo-Hermitian and PT-symmetric quantum mechanics}},}\ }\href {\doibase
  10.1088/0305-4470/37/48/009} {\bibfield  {journal} {\bibinfo  {journal}
  {Journal of Physics A: Mathematical and General}\ }\textbf {\bibinfo {volume}
  {37}},\ \bibinfo {pages} {11645--11679} (\bibinfo {year} {2004})}\BibitemShut
  {NoStop}%
\bibitem [{\citenamefont {Jones}(2005)}]{Jones2005}%
  \BibitemOpen
  \bibfield  {author} {\bibinfo {author} {\bibfnamefont {H~F}\ \bibnamefont
  {Jones}},\ }\bibfield  {title} {\enquote {\bibinfo {title} {{On
  pseudo-Hermitian Hamiltonians and their Hermitian counterparts}},}\ }\href
  {\doibase 10.1088/0305-4470/38/8/010} {\bibfield  {journal} {\bibinfo
  {journal} {Journal of Physics A: Mathematical and General}\ }\textbf
  {\bibinfo {volume} {38}},\ \bibinfo {pages} {1741--1746} (\bibinfo {year}
  {2005})}\BibitemShut {NoStop}%
\bibitem [{\citenamefont {Bender}\ \emph {et~al.}(1999)\citenamefont {Bender},
  \citenamefont {Boettcher},\ and\ \citenamefont {Meisinger}}]{Bender1999}%
  \BibitemOpen
  \bibfield  {author} {\bibinfo {author} {\bibfnamefont {Carl~M.}\ \bibnamefont
  {Bender}}, \bibinfo {author} {\bibfnamefont {Stefan}\ \bibnamefont
  {Boettcher}}, \ and\ \bibinfo {author} {\bibfnamefont {Peter~N.}\
  \bibnamefont {Meisinger}},\ }\bibfield  {title} {\enquote {\bibinfo {title}
  {{$\mathcal{PT}$-symmetric quantum mechanics}},}\ }\href {\doibase
  10.1063/1.532860} {\bibfield  {journal} {\bibinfo  {journal} {Journal of
  Mathematical Physics}\ }\textbf {\bibinfo {volume} {40}},\ \bibinfo {pages}
  {2201--2229} (\bibinfo {year} {1999})}\BibitemShut {NoStop}%
\bibitem [{\citenamefont {Dorey}\ \emph {et~al.}(2001)\citenamefont {Dorey},
  \citenamefont {Dunning},\ and\ \citenamefont {Tateo}}]{Dorey2001}%
  \BibitemOpen
  \bibfield  {author} {\bibinfo {author} {\bibfnamefont {Patrick}\ \bibnamefont
  {Dorey}}, \bibinfo {author} {\bibfnamefont {Clare}\ \bibnamefont {Dunning}},
  \ and\ \bibinfo {author} {\bibfnamefont {Roberto}\ \bibnamefont {Tateo}},\
  }\bibfield  {title} {\enquote {\bibinfo {title} {{Spectral equivalences,
  Bethe ansatz equations, and reality properties in $\mathcal{PT}$-symmetric
  quantum mechanics}},}\ }\href {\doibase 10.1088/0305-4470/34/28/305}
  {\bibfield  {journal} {\bibinfo  {journal} {Journal of Physics A:
  Mathematical and General}\ }\textbf {\bibinfo {volume} {34}},\ \bibinfo
  {pages} {5679--5704} (\bibinfo {year} {2001})}\BibitemShut {NoStop}%
\bibitem [{\citenamefont {Bender}\ \emph {et~al.}(2002)\citenamefont {Bender},
  \citenamefont {Brody},\ and\ \citenamefont {Jones}}]{Bender2002}%
  \BibitemOpen
  \bibfield  {author} {\bibinfo {author} {\bibfnamefont {Carl~M.}\ \bibnamefont
  {Bender}}, \bibinfo {author} {\bibfnamefont {Dorje~C.}\ \bibnamefont
  {Brody}}, \ and\ \bibinfo {author} {\bibfnamefont {Hugh~F.}\ \bibnamefont
  {Jones}},\ }\bibfield  {title} {\enquote {\bibinfo {title} {{Complex
  Extension of Quantum Mechanics}},}\ }\href {\doibase
  10.1103/physrevlett.89.270401} {\bibfield  {journal} {\bibinfo  {journal}
  {Physical Review Letters}\ }\textbf {\bibinfo {volume} {89}},\ \bibinfo
  {pages} {270401} (\bibinfo {year} {2002})}\BibitemShut {NoStop}%
\bibitem [{\citenamefont {Bender}(2007)}]{Bender2007}%
  \BibitemOpen
  \bibfield  {author} {\bibinfo {author} {\bibfnamefont {Carl~M}\ \bibnamefont
  {Bender}},\ }\bibfield  {title} {\enquote {\bibinfo {title} {{Making sense of
  non-Hermitian Hamiltonians}},}\ }\href {\doibase 10.1088/0034-4885/70/6/r03}
  {\bibfield  {journal} {\bibinfo  {journal} {Reports on Progress in Physics}\
  }\textbf {\bibinfo {volume} {70}},\ \bibinfo {pages} {947--1018} (\bibinfo
  {year} {2007})}\BibitemShut {NoStop}%
\bibitem [{\citenamefont {Ashida}\ \emph {et~al.}(2020)\citenamefont {Ashida},
  \citenamefont {Gong},\ and\ \citenamefont {Ueda}}]{Ashida2020}%
  \BibitemOpen
  \bibfield  {author} {\bibinfo {author} {\bibfnamefont {Yuto}\ \bibnamefont
  {Ashida}}, \bibinfo {author} {\bibfnamefont {Zongping}\ \bibnamefont {Gong}},
  \ and\ \bibinfo {author} {\bibfnamefont {Masahito}\ \bibnamefont {Ueda}},\
  }\bibfield  {title} {\enquote {\bibinfo {title} {{Non-Hermitian physics}},}\
  }\href {\doibase 10.1080/00018732.2021.1876991} {\bibfield  {journal}
  {\bibinfo  {journal} {Advances in Physics}\ }\textbf {\bibinfo {volume}
  {69}},\ \bibinfo {pages} {249--435} (\bibinfo {year} {2020})}\BibitemShut
  {NoStop}%
\bibitem [{\citenamefont {Jin}\ and\ \citenamefont {Song}(2010)}]{Jin2010}%
  \BibitemOpen
  \bibfield  {author} {\bibinfo {author} {\bibfnamefont {L.}~\bibnamefont
  {Jin}}\ and\ \bibinfo {author} {\bibfnamefont {Z.}~\bibnamefont {Song}},\
  }\bibfield  {title} {\enquote {\bibinfo {title} {{Physics counterpart of the
  $\mathcal{PT}$ non-Hermitian tight-binding chain}},}\ }\href {\doibase
  10.1103/physreva.81.032109} {\bibfield  {journal} {\bibinfo  {journal}
  {Physical Review A}\ }\textbf {\bibinfo {volume} {81}},\ \bibinfo {pages}
  {032109} (\bibinfo {year} {2010})}\BibitemShut {NoStop}%
\bibitem [{\citenamefont {Zhang}\ \emph {et~al.}(2012)\citenamefont {Zhang},
  \citenamefont {Jin},\ and\ \citenamefont {Song}}]{Zhang2012}%
  \BibitemOpen
  \bibfield  {author} {\bibinfo {author} {\bibfnamefont {X.~Z.}\ \bibnamefont
  {Zhang}}, \bibinfo {author} {\bibfnamefont {L.}~\bibnamefont {Jin}}, \ and\
  \bibinfo {author} {\bibfnamefont {Z.}~\bibnamefont {Song}},\ }\bibfield
  {title} {\enquote {\bibinfo {title} {{Perfect state transfer in
  $\mathcal{PT}$-symmetric non-Hermitian networks}},}\ }\href {\doibase
  10.1103/physreva.85.012106} {\bibfield  {journal} {\bibinfo  {journal}
  {Physical Review A}\ }\textbf {\bibinfo {volume} {85}},\ \bibinfo {pages}
  {012106} (\bibinfo {year} {2012})}\BibitemShut {NoStop}%
\bibitem [{\citenamefont {Zhang}\ \emph {et~al.}(2024)\citenamefont {Zhang},
  \citenamefont {Zhang},\ and\ \citenamefont {Song}}]{ZHP2024}%
  \BibitemOpen
  \bibfield  {author} {\bibinfo {author} {\bibfnamefont {H.~P.}\ \bibnamefont
  {Zhang}}, \bibinfo {author} {\bibfnamefont {K.~L.}\ \bibnamefont {Zhang}}, \
  and\ \bibinfo {author} {\bibfnamefont {Z.}~\bibnamefont {Song}},\ }\bibfield
  {title} {\enquote {\bibinfo {title} {{Dynamics of non-Hermitian Floquet
  Wannier-Stark system}},}\ }\href@noop {} {\  (\bibinfo {year} {2024})},\
  \Eprint {http://arxiv.org/abs/2401.13286} {arXiv:2401.13286 [quant-ph]}
  \BibitemShut {NoStop}%
\bibitem [{\citenamefont {Zhang}\ and\ \citenamefont {Song}(2024)}]{ZKL2024}%
  \BibitemOpen
  \bibfield  {author} {\bibinfo {author} {\bibfnamefont {K.~L.}\ \bibnamefont
  {Zhang}}\ and\ \bibinfo {author} {\bibfnamefont {Z.}~\bibnamefont {Song}},\
  }\bibfield  {title} {\enquote {\bibinfo {title} {Magnetic bloch oscillations
  in a non-hermitian quantum ising chain},}\ }\href@noop {} {\  (\bibinfo
  {year} {2024})},\ \Eprint {http://arxiv.org/abs/2401.17586} {arXiv:2401.17586
  [cond-mat.mes-hall]} \BibitemShut {NoStop}%
\bibitem [{\citenamefont {Maksimov}\ \emph
  {et~al.}(2015{\natexlab{a}})\citenamefont {Maksimov}, \citenamefont
  {Bulgakov},\ and\ \citenamefont {Kolovsky}}]{Maksimov2015}%
  \BibitemOpen
  \bibfield  {author} {\bibinfo {author} {\bibfnamefont {Dmitrii~N.}\
  \bibnamefont {Maksimov}}, \bibinfo {author} {\bibfnamefont {Evgeny~N.}\
  \bibnamefont {Bulgakov}}, \ and\ \bibinfo {author} {\bibfnamefont
  {Andrey~R.}\ \bibnamefont {Kolovsky}},\ }\bibfield  {title} {\enquote
  {\bibinfo {title} {{Wannier-Stark states in double-periodic lattices. I.
  One-dimensional lattices}},}\ }\href {\doibase 10.1103/physreva.91.053631}
  {\bibfield  {journal} {\bibinfo  {journal} {Physical Review A}\ }\textbf
  {\bibinfo {volume} {91}},\ \bibinfo {pages} {053631} (\bibinfo {year}
  {2015}{\natexlab{a}})}\BibitemShut {NoStop}%
\bibitem [{\citenamefont {Maksimov}\ \emph
  {et~al.}(2015{\natexlab{b}})\citenamefont {Maksimov}, \citenamefont
  {Bulgakov},\ and\ \citenamefont {Kolovsky}}]{Maksimov2015a}%
  \BibitemOpen
  \bibfield  {author} {\bibinfo {author} {\bibfnamefont {Dmitrii~N.}\
  \bibnamefont {Maksimov}}, \bibinfo {author} {\bibfnamefont {Evgeny~N.}\
  \bibnamefont {Bulgakov}}, \ and\ \bibinfo {author} {\bibfnamefont
  {Andrey~R.}\ \bibnamefont {Kolovsky}},\ }\bibfield  {title} {\enquote
  {\bibinfo {title} {{Wannier-Stark states in double-periodic lattices. II.
  Two-dimensional lattices}},}\ }\href {\doibase 10.1103/physreva.91.053632}
  {\bibfield  {journal} {\bibinfo  {journal} {Physical Review A}\ }\textbf
  {\bibinfo {volume} {91}},\ \bibinfo {pages} {053632} (\bibinfo {year}
  {2015}{\natexlab{b}})}\BibitemShut {NoStop}%
\bibitem [{\citenamefont {Corrielli}\ \emph {et~al.}(2013)\citenamefont
  {Corrielli}, \citenamefont {Crespi}, \citenamefont {Della~Valle},
  \citenamefont {Longhi},\ and\ \citenamefont {Osellame}}]{Corrielli2013}%
  \BibitemOpen
  \bibfield  {author} {\bibinfo {author} {\bibfnamefont {Giacomo}\ \bibnamefont
  {Corrielli}}, \bibinfo {author} {\bibfnamefont {Andrea}\ \bibnamefont
  {Crespi}}, \bibinfo {author} {\bibfnamefont {Giuseppe}\ \bibnamefont
  {Della~Valle}}, \bibinfo {author} {\bibfnamefont {Stefano}\ \bibnamefont
  {Longhi}}, \ and\ \bibinfo {author} {\bibfnamefont {Roberto}\ \bibnamefont
  {Osellame}},\ }\bibfield  {title} {\enquote {\bibinfo {title} {{Fractional
  Bloch oscillations in photonic lattices}},}\ }\href {\doibase
  10.1038/ncomms2578} {\bibfield  {journal} {\bibinfo  {journal} {Nature
  Communications}\ }\textbf {\bibinfo {volume} {4}} (\bibinfo {year} {2013}),\
  10.1038/ncomms2578}\BibitemShut {NoStop}%
\bibitem [{\citenamefont {Longhi}(2011)}]{Longhi2011}%
  \BibitemOpen
  \bibfield  {author} {\bibinfo {author} {\bibfnamefont {Stefano}\ \bibnamefont
  {Longhi}},\ }\bibfield  {title} {\enquote {\bibinfo {title} {{Photonic Bloch
  oscillations of correlated particles}},}\ }\href {\doibase
  10.1364/ol.36.003248} {\bibfield  {journal} {\bibinfo  {journal} {Optics
  Letters}\ }\textbf {\bibinfo {volume} {36}},\ \bibinfo {pages} {3248}
  (\bibinfo {year} {2011})}\BibitemShut {NoStop}%
\bibitem [{\citenamefont {Christodoulides}\ \emph {et~al.}(2003)\citenamefont
  {Christodoulides}, \citenamefont {Lederer},\ and\ \citenamefont
  {Silberberg}}]{Christodoulides2003}%
  \BibitemOpen
  \bibfield  {author} {\bibinfo {author} {\bibfnamefont {Demetrios~N.}\
  \bibnamefont {Christodoulides}}, \bibinfo {author} {\bibfnamefont {Falk}\
  \bibnamefont {Lederer}}, \ and\ \bibinfo {author} {\bibfnamefont {Yaron}\
  \bibnamefont {Silberberg}},\ }\bibfield  {title} {\enquote {\bibinfo {title}
  {{Discretizing light behaviour in linear and nonlinear waveguide
  lattices}},}\ }\href {\doibase 10.1038/nature01936} {\bibfield  {journal}
  {\bibinfo  {journal} {Nature}\ }\textbf {\bibinfo {volume} {424}},\ \bibinfo
  {pages} {817--823} (\bibinfo {year} {2003})}\BibitemShut {NoStop}%
\end{thebibliography}%
	
\end{document}